\documentclass[a4paper,prd,noshowpacs,noshowkeys,notitlepage,nofootinbib]{revtex4}
\usepackage[T1]{fontenc}
\usepackage{ae,aecompl} 
\usepackage{amsmath,amssymb,mathrsfs,amsthm}
\usepackage{bbold}
\usepackage{blkarray}
\usepackage{mathtools,slashed}
\usepackage{graphicx}
\usepackage[]{units}
\usepackage[usenames,dvipsnames,svgnames]{xcolor} 
\usepackage{hyperref}
\hypersetup{
   colorlinks=true,       
    linkcolor=Blue,          
    citecolor=Plum,        
    filecolor=magenta,      
    urlcolor=YellowOrange           
}
\usepackage{ccnishi}
\usepackage[normalem]{ulem} 
\usepackage{setspace}  
\usepackage{tikz-feynman}
\tikzfeynmanset{compat=1.1.0}
\usepackage{xfrac} 
\makeatletter
\DeclareRobustCommand{\cev}[1]{%
  {\mathpalette\do@cev{#1}}%
}
\newcommand{\do@cev}[2]{%
  \vbox{\offinterlineskip
    \sbox\z@{$\m@th#1 x$}%
    \ialign{##\cr
      \hidewidth\reflectbox{$\m@th#1\vec{}\mkern4mu$}\hidewidth\cr
      \noalign{\kern-\ht\z@}
      $\m@th#1#2$\cr
    }%
  }%
}
\makeatother

\makeatletter
\newcommand{\overleftrightsmallarrow}{\mathpalette{\overarrowsmall@\leftrightarrowfill@}}
\newcommand{\overrightsmallarrow}{\mathpalette{\overarrowsmall@\rightarrowfill@}}
\newcommand{\overleftsmallarrow}{\mathpalette{\overarrowsmall@\leftarrowfill@}}
\newcommand{\overarrowsmall@}[3]{%
  \vbox{%
    \ialign{%
      ##\crcr
      #1{\smaller@style{#2}}\crcr
      \noalign{\nointerlineskip}%
      $\m@th\hfil#2#3\hfil$\crcr
    }%
  }%
}
\def\smaller@style#1{%
  \ifx#1\displaystyle\scriptstyle\else
    \ifx#1\textstyle\scriptstyle\else
      \scriptscriptstyle
    \fi
  \fi
}
\makeatother
\newcommand{\oLR}[1]{\overleftrightsmallarrow{#1}} 
\DeclareMathOperator{\STr}{STr}

\providecommand{\deldel}[2]{\frac{\delta #1}{\delta #2}}

\providecommand{\cQ}{\mathcal{Q}}
\providecommand{\cM}{\mathscr{M}}

\providecommand{\cY}{\mathscr{Y}}
\providecommand{\cF}{\mathscr{F}}
\providecommand{\cO}{\mathcal{O}}

\providecommand{\la}[1]{\lambda_{#1}}

\providecommand{\tL}{\tilde{L}}
\providecommand{\td}{\tilde{d}}

\providecommand{\tw}{\tilde{w}}

\providecommand{\btheta}{\bar{\theta}}
\providecommand{\Lcp}{\Lambda_{\rm cp}}
\providecommand{\Mvlq}{M_{\rm VLQ}}
\providecommand{\Mpl}{M_{\rm Pl}}

\providecommand{\cp}{\mathsf{CP}}

\providecommand{\hY}{\hat{Y}}
\providecommand{\hU}{\hat{U}}

\providecommand{\lag}{\mathscr{L}}

\providecommand{\eps}{\epsilon}

\providecommand{\Str}{\operatorname{Str}}

\providecommand{\keff}{\kappa_{\rm eff}}

\begin{document}
\title{
Effective description of Nelson-Barr models and the theta parameter
}
\author{Gustavo~H.~S.~Alves}
\email{alves.gustavo@ufabc.edu.br}
\affiliation{Centro de Ci\^{e}ncias Naturais e Humanas,\\
Universidade Federal do ABC, 09.210-170,
Santo Andr\'{e}-SP, Brasil}
\author{Celso~C.~Nishi}
\email{celso.nishi@ufabc.edu.br}
\affiliation{
Centro de Matemática, Computação e Cognição\\
Universidade Federal do ABC -- UFABC, 09.210-170,
Santo André, SP, Brazil
}
\begin{abstract}
Nelson-Barr theories solve the strong CP problem with CP spontaneously broken at $\Lcp$ and transmitted to the CP 
conserving version of the SM through vector-like quarks (VLQs).
For an arbitrary number of CP breaking scalars and VLQs, we perform the full one-loop matching calculations at $\Lcp$ 
up to dimension five operators and the relevant matching calculations at the VLQ scale relevant to tracking the 
contributions to the parameter $\btheta$.
In the EFT after the CP breaking scalars have been integrated out, we confirm that the running of $\btheta$ at 
\emph{one-loop} is induced by CP violating dimension five operators and similarly by dimension six operator 
after matching to the SMEFT.
In the latter, an additional contribution enters in the matching to $\btheta$ at \emph{tree-level} due to a dimension 
five operator.
Analyzing the experimental constraint from $\btheta$ for generic settings, 
we see that a large separation between $\Lcp$ and the VLQ scale already suppresses the one-loop 
contribution to $\btheta$ sufficiently.
We confirm that a simple extension based on a nonconventional CP has vanishing one-loop contribution to $\btheta$.
Part of our results can be also applied to generic VLQ extensions of the SM.
\end{abstract}
\maketitle
\section{Introduction}

The only known source of CP violation in the SM is a single order one irremovable CKM 
phase\,\cite{ParticleDataGroup:2022pth} residing in
the Yukawa couplings between quarks and the Higgs, which manifests exclusively in flavor violating processes. The
measurement of a similar phase in the leptonic sector is one of the major goals of planned neutrino oscillation
experiments. In contrast, although a flavor blind source of CP violation exists in the SM in the form of the 
$\btheta$
term of QCD, the nonobservation of the electric dipole moment of the neutron constrains this parameter to be tiny:
$\btheta\lesssim 10^{-10}$\,\cite{nEDM:exp}. This naturality problem ---known as the strong CP 
problem\,\cite{Hook:2018dlk,EDMreview}--- 
becomes more evident when we consider that two sources in $\btheta$ should cancel:
\[
\btheta = \theta - \arg\det(\mathcal{M})\,,
\]
the first is intrinsic to the nontrivial vacuum of QCD while $\mathcal{M}$ collects the quark mass matrices that carry 
the order one CKM phase.

Solving this problem without invoking an axion involves the promotion of CP\,\cite{strongCP:CP,nelson,barr} or 
P\,\cite{strongCP:P}\,\footnote{See Ref.\,\cite{craig:P} for a recent analysis.} 
as a fundamental symmetry of nature that is spontaneously broken.
The challenge is to arrange this breaking to generate an order one CKM phase still suppressing $\btheta$ 
sufficiently.
The most well known example involving CP is based on the Nelson-Barr\,\cite{nelson,barr} idea which guarantees 
vanishing $\btheta$ at tree-level by allowing CP breaking only in the mixing between SM quarks and heavy vector-like 
quarks (VLQs), also denoted as VLQs of Nelson-Barr type (NB-VLQs) in Refs.\,\cite{nb-vlq,nb-vlq:fit,nb-vlq:more}.
Different alternatives with spontaneous CP include the use of non-renormalization theorems of 
supersymmetry\,\cite{hiller:01}, modular flavor symmetries\,\cite{strongcp:modular}, different structures or symmetries 
to ensure texture-zeros in the quark mass matrices\,\cite{texturezero:det} or Yukawa matrices in multi-Higgs extensions 
\cite{strongcp:multihiggs}.

The challenge of the axionless solutions such as the Nelson-Barr setting to naturally suppress higher order 
corrections to $\btheta$ is inexistent within the SM:
the running of $\btheta$ is expected to arise only at seven loops\,\cite{ellis.gaillard,vainshtein} and finite threshold 
corrections occur at four loops \cite{Khriplovich:1985jr}.
So the model building challenge for the UV extension is to suppress the corrections beyond the SM,
which may already show up at one-loop in Nelson-Barr extensions\,\cite{nelson,dine} and is also generically expected in 
other theories\,\cite{vecchi:theta}.
The one-loop correction can be confirmed in the simplest NB implementation, the BBP model\,\cite{BBP}, which only 
adds one complex scalar 
and one VLQ to the SM.
By using a nonconventional CP symmetry in a model denoted as the CP4 model\,\cite{cp4}, these corrections may be 
postponed to two-loops.\footnote{
In the minimal left-right extension of the SM, it can be postponed to three-loops\,\cite{LR:theta}.
}

Nevertheless, the typical corrections to $\btheta$ that appear at one- or two-loops are not so problematic because they 
can be made 
under control if we suppress the couplings of the CP breaking scalars to these NB-VLQs or to the SM Higgs.\footnote{
For the latter, it is assumed that the hierarchy problem is solved by other means.
For example, the two problems can be connected in Nelson-Barr relaxion models\,\cite{Davidi:2017gir}
or using SUSY\,\cite{dine}.
}
There are, however, corrections at three-loops that cannot be arbitrarily suppressed because they depend only on the 
Yukawa couplings of NB-VLQs to the SM\,\cite{vecchi.1}, the same ones sourcing the CP violation. 
These irreducible three-loop contributions to $\btheta$ were analyzed through the construction of CP odd 
flavor invariants which are now completely mapped out in terms of 9 CP odd invariants in the 
generating set of 29 invariants\,\cite{vlq.hilbert}.

Concerning the CP breaking scalars, as they generate a significant part of the mass of the NB-VLQs, they should 
lie above their scale which in turn  are constrained above the TeV scale from collider searches.
On the other hand, the stability of the domain walls from the spontaneous breaking of the exact CP 
symmetry\,\cite{reece:stable}, requires that this breaking should occur before inflation and the CP 
symmetry cannot be restored in the reheating phase. This leads to 
a number of cosmological implications\,\cite{reece}.
However, raising $\Lcp$ too much worsens the Nelson-Barr quality problem\,\cite{dine,reece,choi.kaplan}: 
gravity induced Planck suppressed operators may generate dimension five operators that break CP with the same CP 
breaking scalars. This leads to an approximate upper limit for $\Lcp$ of around $10^8\,\unit{GeV}$.
Additional gauge symmetries\,\cite{reece}, supersymmetry\,\cite{dine,evans.yokozaki}, strong 
dynamics\,\cite{vecchi.2} or approximate global symmetries\,\cite{murai.nakayama:1}
can improve the quality and allow for higher values of $\Lcp$.

Therefore, we focus here on the scenario where the NB-VLQs lie at the TeV scale while the CP breaking 
sector is pushed to high energy.
This is the scenario we considered in Refs.\,\cite{nb-vlq,nb-vlq:fit} and we have studied the phenomenological 
constraints coming from direct renormalizable Yukawa interactions between VLQs, SM quarks and Higgs.
There we have found through explicit parametrization that these VLQs typically couple to the SM quarks and Higgs 
following the hierarchy of the CKM last row or column, a feature that alleviates the strongest flavor constraints that 
apply to the first two quark families.
We have found that it is possible to distinguish NB-VLQs from usual VLQs in restricted regions of parameter space but 
generically the distinction will be difficult.

Here we intend to study the additional effects induced by higher dimensional operators beyond the renormalizable 
interactions using an Effective Field Theory (EFT) framework. 
As we assume the CP breaking scale $\Lcp$ is high while the NB-VLQs are at the TeV scale, we need to perform a 
two-stage matching procedure: integrating out the CP breaking scalars at $\Lcp$ and then integrating out the NB-VLQs at 
the TeV scale.
This procedure will leave imprints in the Standard Model Effective Field Theory (SMEFT) 
dimension six operators\,\cite{buchmuller.wyler,warsaw}.
The EFT calculation also has the advantage that some partial results may be applied to the more general context of the 
SM augmented with VLQs.
The methods to perform the top-down matching from a UV theory to the SMEFT has undergone substantial theoretical 
developments in recent years, particularly through the use of functional matching techniques \cite{cohen.zhang:20,
HLM:1,Henning:2016lyp, Drozd:2015rsp,Fuentes-Martin:2016uol} which was built upon previous works
\cite{Cheyette:1987qz,Chan:1986jq,Gaillard:1985uh} that led to the Covariant Derivative Expansion (CDE) in which 
manifestly gauge covariant results can be directly obtained.
This leads to several advantages compared\,\cite{Cohen:2022tir} to diagrammatic techniques.
The one-loop case can be now fully automated in package such as \texttt{STrEAM} \cite{Cohen:2020qvb} and 
\texttt{Matchete} \cite{matchete} implemented in Wolfram \texttt{Mathematica}.
They have been successfully applied to a variety of physics scenarios beyond the SM  \cite{Cohen:2022tir}.

The outline of this article is as follows:
In Sec.\,\ref{sec:NB}, we review the model of singlet NB-VLQs  and establish our notation.
Section~\ref{sec:func.match} briefly reviews the one-loop matching procedure using the functional method and
CDE.
Sections \ref{sec:tree:Lcp} and \ref{sec:1L:Lcp} deal respectively with tree and one-loop matching when we integrate 
out the CP breaking scalars of the theory.
The one-loop matching is carried out in full up to dimension five operators
and the resulting theory is an EFT containing VLQs interacting with renormalizable as well as dimension five 
interactions.
We denote this theory as the VLQ-EFT.
In section \ref{sec:Mvlq} we perform the matching when we integrate out the VLQs from the previous VLQ-EFT.
The one-loop matching is performed focusing on the contribution to $\btheta$.
We calculate the non-trivial one-loop RGE for $\btheta$ induced by the CP violating dimension five operators.
We also confirm that $\btheta$ still runs at one-loop within the SMEFT and clarify a subtlety in the matching  
of the VLQ-EFT to the SMEFT where the QCD $\theta$ term should receive a matching contribution at \emph{tree-level}.
Some of the results can be applied more broadly to the VLQ-EFT that does not need to descend from a Nelson-Barr theory.
Section \ref{sec:results} presents the final formulas, generic analysis of the one-loop contribution to $\btheta$
and their application to two simple NB models:
the CP4 model and the BBP model.
We also discuss briefly the constraints coming from CP violating operators to the quark dipole and 
chromo-electric moments.
Readers interested only in the results should skip directly to Sec.\,\ref{sec:results}.
Finally, the summary is presented in Sec.\,\ref{sec:summary}.

\section{Nelson-Barr theories}
\label{sec:NB}

Nelson-Barr theories\,\cite{nelson,barr} are characterized by the presence of a CP breaking sector, usually composed of 
scalars, and vector-like quarks (VLQs) which transmit the CP breaking to the CP conserving version of the SM.
Here we assume these VLQs are down-type singlet VLQs $B_{aL},B_{bR}$, $a,b=1,\dots,n_B$, with the same quantum numbers 
as the SM down-type singlet quark $d_{pR}$, $p=1,2,3$.
Up-type singlet VLQs may be equally considered.
We can assume a generic CP breaking sector composed of real singlet scalars $s_i$.
Assuming no other degree of freedom up to the CP breaking scale, we can write the Lagrangian in the CP broken phase as 
\eqali{
\label{lag:UV:cp}
\lag_{\rm UV}&=\text{(kinetic)}
-\Big(\bar{q}_{L}\cY^u\tilde{H}u_{R}+\bar{q}_L\cY^d Hd_{R}+h.c.\Big)
\cr
&\quad 
-\Big(\bar{B}_{L}\cM^{Bd}d_{R}
+\bar{B}_L(\cF_is_i)d_R
+\bar{B}_{L}\cM^BB_{R}
+h.c.\Big)
-V(|H|^2,s_i)
\,,
}
where $q_L,u_R,d_R$ are the SM quark fields and $H$ is the Higgs doublet.
In flavor space, we adopt a matrix notation where, e.g., $\bar{q}_L\cY^d Hd_{R}=\bar{q}_{pL}[\cY^d]^{pr} Hd_{rR}$ and 
$\bar{B}_{L}\cM^{Bd}d_{R}=\bar{B}_{aL}[\cM^{Bd}]^{ar}d_{rR}$, so that $\cY^d$ is a $3\times 3$ matrix and $\cM^{Bd}$ is 
a $n_B\times 3$ matrix.
When not specified, we use the basis where $\cY^u=\hat{\cY}^u$ is diagonal, with a hat denoting diagonal matrices.

In the Nelson-Barr setting, with conventional\footnote{%
In Sec.\,\ref{sec:cp4} we review a model with non-conventional CP.}
CP,
$\cY^u,\cY^d$ are \textit{real} $3\times 3$ matrices, $\cM^B$ is a real bare mass matrix and only $\cM^{Bd}$ is a 
complex matrix induced by the CP breaking vevs $u_i$:
\eq{
\label{MBd:ui}
\cM^{Bd}=\cF_iu_i\,.
}
Prior to spontaneous CP breaking, $s_i+u_i$ are the fundamental scalars which can be CP even or odd and their 
properties 
restrict the couplings $\cF_i$. 
For example, in the minimal BBP model\,\cite{BBP}, there are only two scalars for which $\cF_1$ is real while $\cF_2$ 
is purely imaginary.
The real couplings and the forbidden terms follow from CP conservation and a $\ZZ_2$ symmetry\,\footnote{A larger 
$\ZZ_n$ or $U(1)$ are also possible\,\cite{dine}.}
under which only $B_{L,R}$ and $s_i+u_i$ are odd.

The CP breaking scale typically lies at a very high scale $\Lcp$.
We assume that below or much below this CP breaking scale, the only BSM states are the VLQs.
The scenario where these VLQs lie close to the TeV scale was studied in Refs.\,\cite{nb-vlq,nb-vlq:fit,nb-vlq:more} 
where they were denoted as VLQs of Nelson-Barr type (NB-VLQs).
Up to dimension four in the Lagrangian, only $\cM^{Bd}$ breaks CP and $\ZZ_2$ softly (spontaneously) realizing the 
Nelson-Barr mechanism that guarantees $\bar{\theta}=0$ at tree level.
Describing NB-VLQs requires one less parameter than generic VLQs which, albeit difficult, leads to  falsefiable 
correlations.

The scalar potential $V$ includes the Higgs potential of the SM adjoined with a generic renormalizable potential of the 
real scalar fields $s_i$:
\eq{\label{VS:general}
V_S=\ums{2}M^2_{ij}s_is_j+\ums{3!}\lambda'_{ijk}s_is_js_k+\ums{4!}\lambda_{ijkl}s_is_js_ks_l\,,
}
and 
\eq{\label{VHS:general}
V_{HS}=H^\dag H(\ums{2}\gamma_{ij}s_is_j+\gamma'_is_i)\,.
}
The summation convention is implicit here and throughout. The summation symbol will be written explicitly when 
ambiguity 
arises.
In the Nelson-Barr setting, $\gamma'_i$ comes from the CP breaking vevs $u_i$ as $\gamma_i'= 
\gamma_{ij}u_j\sim \gamma_{ij}\Lcp$.
For the Higgs potential we will use for the quadratic term,
\eq{
V_H\supset \mu_H^2|H|^2\,.
}
Note that the parameter $\mu_H^2$ differs from the one in the SM by threshold corrections at $\Lcp$ which may be 
as large as 
$\Lcp^2/16\pi^2$. This is reminiscent of the hierarchy problem and needs to be solved by other means.

In flavor space (of SM and VLQs), the Lagrangian \eqref{lag:UV:cp} is written in the basis where CP symmetry and the 
possible additional symmetry such as $\ZZ_2$ is manifest.
We denote that basis as the \emph{CP basis} of the UV theory.
The VLQs, however, have no definite mass after spontaneous CP breaking due to the mixing term $\cM^{Bd}$.
We can go to the \emph{VLQ mass basis} with Lagrangian
\eqali{
\label{lag:UV:mass}
\lag_{\rm UV}&=\text{(kinetic)}
    -\Big(\bar{q}_{L}Y^u\tilde{H}u_{R}+\bar{q}_LY^d Hd_{R}
    +\bar{q}_LY^B HB_{R}+h.c.\Big)
\cr
&\quad 
-\Big(\bar{B}_{L}M^BB_{R}
    +\bar{B}_L(F_is_i)d_R
    +\bar{B}_L(G_is_i)B_R
    +h.c.\Big)
-V(|H|^2,s_i)
\,.
}
Here we use roman letters to denote the couplings and masses in contrast to calligraphic letters used in the CP basis 
\eqref{lag:UV:cp}.
We continue to adopt a matrix notation in the flavor indices where, e.g., 
$\bar{B}_LF_id_R=\bar{B}_{aL}[F_i]^{ap}d_{pR}$.
Usually, we assume the basis where $Y^u=\hY^u$ is diagonal.

The change of basis from \eqref{lag:UV:cp} to \eqref{lag:UV:mass}
can be performed by an analytic rotation only in the space $(d_R,B_R)$\,\cite{nb-vlq:more}:
\eq{
\label{WR:w}
\mtrx{d_R\cr B_R}\to 
\mtrx{\big(\id_3-ww^\dag\big)^{1/2} & w\cr -w^\dag & \big(\id_n-w^\dag w\big)^{1/2}}
\mtrx{d_R\cr B_R}
\,,
}
where $n=n_B$ in $\id_n$.
The transformation matrix can be also written as
\eq{
W_R=\mtrx{\id_3 & \tw \cr -\tw^\dag & \id_n}
\mtrx{(\id_3+\tw\tw^\dag)^{-1/2} & 0 \cr 0 & (\id_n+\tw^\dag \tw)^{-1/2}}
\,,
}
One simple choice leads to
\subeqali[WR:YdYB]{
\label{Yd:NB:1/2}
Y^d&= \cY^d(\id_3+\tw\tw^\dag)^{-1/2}=\cY^d(\id_3-ww^\dag)^{+1/2}\,,
\\
\label{YB:cal-Yd}
Y^B&= \cY^d \tw(\id_n+\tw^\dag \tw)^{-1/2}=\cY^dw\,,
\\
\label{WR:MB}
M^B&=\cM^B(\id_n+\tw^\dag \tw)^{+1/2}=\cM^B(\id_n-w^\dag w)^{-1/2}\,,
\\
F_i&=\cF_i(\id_3+\tw\tw^\dag)^{-1/2}=\cF_i(\id_3-ww^\dag)^{+1/2}\,,
\\
G_i&=\cF_i\tw(\id_n+\tw^\dag \tw)^{-1/2}=\cF_i w\,,
}
where 
\eq{
\label{def:w}
w=\tw(\id_n+\tw^\dag\tw)^{-1/2}\,,\quad
\tw^\dag = {\cM^B}^{-1}\cM^{Bd}\,.
}
We can also write in implicit form,
\eq{
\label{wdagger}
w^\dag={M^B}^{-1}{\cM^{Bd}}\,.
}
Note that $M^B$ in \eqref{WR:MB} is not generic as ${\cM^B}^{-1}M^B$ must be hermitean.

For a generic VLQ, the Yukawa couplings $Y^d$ and $Y^B$ would be independent but in the Nelson-Barr setting they are 
related by 
the very interesting relation\,\cite{nb-vlq,nb-vlq:more}:
\eq{
\label{YB:Yd}
Y^B=Y^d\tw\,,
}
which shows that roughly $Y^B$ inherits the hierarchy of SM Yukawas.
In fact, the inherited hierarchy is typically from the CKM third column:
\eq{
\label{typical.YB}
|Y^B_1|:|Y^B_2|:|Y^B_3|\sim |V_{ub}|:|V_{cb}|:|V_{tb}|
\sim 0.0036:0.04:1
\,,
}
where $|Y^B_i|$ is the norm of $Y^B_{ia}$, $a=1,\dots,n$.
The typical hierarchy for up-type NB-VLQs would follow the third \emph{row} of the CKM.
Analogously to \eqref{YB:Yd}, the Yukawa couplings $F_i$ and $G_i$ are also related by 
\eq{
\label{Gi:Fi}
G_i=F_i\tw\,.
}
Given \eqref{YB:Yd}, this relation can be also fully written in terms of the couplings in the VLQ mass basis as
\eq{
\label{Gi:Fi.YB}
G_i=F_i{Y^d}^{-1}Y^B\,.
}
These Yukawa couplings $F_i$ and $G_i$ are also related to the mass matrices through the vevs $u_i$; cf. \eqref{MBd:ui}.
Some of the relations that can be written are
\eqali{
\label{Gi.ui}
F_iu_i&=\cM^Bw^\dag\,,\cr
G_iu_i&=M^Bw^\dag w\,.
}

Our goal here is to calculate the one-loop contributions to $\btheta$ in Nelson-Barr theories.
For that end, there are two scales of interest:
the CP breaking scale $\Lcp$ where we integrate out the scalars $s_i$ and the scale $\Mvlq$ where we integrate out the 
NB-VLQs. After this last step, we get the SMEFT.
Schematically we have the following tower of theories:
\eq{
\label{efts}
\text{Nelson-Barr theory} ~~\stackrel{\Lcp}{\longleftrightarrow}~~
\text{SM + NB-VLQs} ~~\stackrel{\Mvlq}{\longleftrightarrow}~~
\text{SMEFT}.
}

\section{Functional matching}
\label{sec:func.match}

One of the key concepts in constructing effective theories is matching. The principle behind matching is that both the 
full theory and the EFT should describe the same physics at the matching scale. To achieve this, we determine the 
Wilson 
coefficients of the EFT as functions of the fundamental parameters of the higher-energy theory, ensuring that the 
observable results of both theories align on the matching scale \cite{Cohen:2022tir}.

In this work, we will use functional matching to determine both the contributions to the effective theory at the 
tree level and at the one-loop. 
The method at one-loop makes use of the recently developed covariant derivative expansion (CDE) 
method\,\cite{HLM:1,Henning:2016lyp} where 
manifestly gauge covariant results are automatic.
We will follow the method of Ref.\,\cite{cohen.zhang:20} closely.

We review the method using two (sets of) scalar fields: a light field $\phi$ and a heavy field $\Phi$.
We first split the fields into the classical 
background configurations, $\Phi_{\mathrm{B}}$ and $\phi_{\mathrm{B}}$, and their quantum fluctuations $\Phi^{\prime}$ 
and $\phi^{\prime}$ as
\begin{equation}
\label{phiBphiprime}
     \Phi=\Phi_{\mathrm{B}}+\Phi^{\prime}, \quad \phi=\phi_{\mathrm{B}}+\phi^{\prime}\,.
\end{equation}
The condition for matching can be stated as
\begin{equation}\label{condicaogamma}
     \Gamma_{\text{L,UV}}\left[\phi_{\mathrm{B}}\right]= \Gamma_{\text{EFT}}\left[\phi_{\mathrm{B}}\right]\,,
\end{equation}
where $\Gamma_{ \mathrm{L,UV}}\left[\phi_{\mathrm{B}}\right] $ is the one-light-particle-irreducible (1LPI) effective 
action of the UV theory that only includes the lights fields
\begin{equation}\label{Gamma UV}
    \begin{aligned}
\Gamma_{ \mathrm{L,UV}}\left[\phi_{\mathrm{B}}\right] & \equiv 
-i\log Z_{\mathrm{UV}}\left[J=0,j\right]-\int \mathrm{d}^{d} x j \phi_{\mathrm{B}}
\,, \\
& \simeq \int \mathrm{d}^{d} x \lag_{\mathrm{UV}}\left(\Phi_{\mathrm{c}}\left[\phi_{\mathrm{B}}\right], 
\phi_{\mathrm{B}}\right)+\frac{i}{2} \log \operatorname{det} 
\cQ_{\mathrm{UV}}\left[\Phi_{\mathrm{c}}\left[\phi_{\mathrm{B}}\right], \phi_{\mathrm{B}}\right],
\end{aligned}
\end{equation}
where the last expansion is truncated at one-loop.
The functional $-i\log Z_{\rm UV}[J,j]$ is the generating functional for the connected $n$-point functions 
(integrated in $\Phi',\phi'$) 
while the classical solution $\Phi_c$ for $\Phi$ in the absence of the source $J$ is given by
\eq{
\label{def:Hc}
\deldel{}{J(x)}\Big(-i\log Z_{\rm UV}[j,J]\Big)\bigg|_{J=0}=\Phi_c(x)=\Phi_c[\phi_B](x)\,.
}
The functional $\cQ_{\rm UV}$ is the quadratic part of the classical UV action.
Analogously, the effective one-particle-irreducible (1PI) action of the effective theory is given by
\begin{equation}\label{Gamma EFT}
    \begin{aligned}
\Gamma_{\mathrm{EFT}}\left[\phi_{\mathrm{B}}\right] & =-i  \log Z_{\mathrm{EFT}}\left[j\right]-\int 
\mathrm{d}^{d} x j \phi_{\mathrm{B}} 
\,,
\\
& \simeq \int \mathrm{d}^{d} 
x\left(\lag_{\mathrm{EFT}}^{\text{tree}}\left(\phi_{\mathrm{B}}\right)+\lag_{\mathrm{EFT}}^{\text{1-loop 
}}\left(\phi_{\mathrm{B}}\right)\right)+\frac{i}{2} \log \operatorname{det} \cQ_{\mathrm{EFT}}.
\end{aligned}
\end{equation}

The matching condition \eqref{condicaogamma} at tree-level thus gives
\begin{equation}
    \lag^{\text{tree}}_{\text{EFT}}(\phi_{\mathrm{B}})=\hat{\lag}_{\text{UV}}(\phi_{\mathrm{B}}, 
\Phi_c[\phi_{\mathrm{B}}])\,,
\end{equation}
where the hat denotes the expansion of $\Phi_c$ in powers $1/M$ of the heavy mass $M$.
In practical terms, the definition of the classical field $\Phi_c$ in \eqref{def:Hc} is the solution of the 
classical equations of motion (EOM) of the heavy field. 

At the next order, the matching condition at one-loop gives
\begin{equation}
\label{matching:1L}
\int \mathrm{d}^{d} x \lag_{\mathrm{EFT}}^{\text{1-loop }}\left[\phi_{\mathrm{B}}\right]+\frac{i}{2} \log 
\operatorname{det} \cQ_{\mathrm{EFT}}\left[\phi_{\mathrm{B}}\right]=\frac{i}{2} \log \operatorname{det} 
\cQ_{\mathrm{UV}}\left[\Phi_c\left[\phi_{\mathrm{B}}\right], \phi_{\mathrm{B}}\right],
\end{equation}
which is solved by
\begin{equation}
\label{del L d phi2}
    \int d^dx\lag^{\text {1-loop}}_{\text{EFT}}=\frac{i}{2}\STr\log 
    \left(
    \cQ_{\mathrm{UV}}\left[\Phi_c\left[\phi_{\mathrm{B}}\right], \phi_{\mathrm{B}}\right]
\right)\Big|_{\text{hard}}.
\end{equation}
The supertrace $\STr$ indicates all the traces in functional space, including the fermionic nature when necessary,
and other spaces such as gauge and flavor.
The cancellation of the light loops in both sides of \eqref{matching:1L} is correctly taken into account by the 
``hard'' expansion where all the light masses are expanded out in the loop integrals using the method of 
regions\,\cite{fuentes-martin}.

For further calculations, it is convenient to decompose
\eq{ \label{K-X}
-\cQ_{\mathrm{UV}}\left[\Phi_c\left[\phi_{\mathrm{B}}\right], \phi_{\mathrm{B}}\right]
=
\frac{\delta \lag_{\rm UV}}{\delta\phi^2}\bigg|_{\Phi=\Phi_c[\phi]}=\mathbf{K}-\mathbf{X},
}
where $\mathbf{K}$ is defined by the ``inverse propagator'' operator of the heavy fields and $\mathbf{X}$ is the 
interaction part.
Usually, we can write the kinetic and mass terms in relativistic theories in block-diagonal form as
\begin{equation}
\lag_{\rm UV}\supset \frac{1}{2}\bar{\varphi}\mathbf{K}\varphi=\frac{1}{2}\sum_i\bar{\varphi}_iK_i\varphi_i\,.
\end{equation}
For each type of field, $K_i$ is equal to
\eqali{
K_i=\begin{cases}
P^2-M^2_i\,,& \text{for scalar},\\
\slashed{P}-M_i\,,&\text{for fermionic},\\ 
-\eta_{\mu\nu}(P^2-M^2_i)+(1-\frac{1}{\xi})P^\mu P^\nu\,,& 
\text{for vector fields}\,,
\end{cases}
}
where $P_\mu\equiv iD_\mu=i\partial_\mu-gA_\mu$ is the Hermitean version of the covariant dervivative.
In turn, the interaction matrix is given by\,\cite{cohen.zhang:20}
\begin{equation}
 \mathbf{X}[\phi,P_\mu]= \mathbf{U}[\phi]+\left(P_\mu \mathbf{Z}[\phi]+ \bar{\mathbf{Z}}[\phi]P_\mu\right)+\cdots
\end{equation}
where $\mathbf{U}[\phi]$, $\mathbf{Z}[\phi]$ and $\bar{\mathbf{Z}}[\phi]$ are matrices that depend on the light fields 
of the theory and the derivatives that act only on these fields.\footnote{%
The derivatives that act only on the field, i.e., $(D_\mu\phi)$, are defined as closed derivatives.
} 
In turn, the factors of $P_\mu$ that multiply $\mathbf{Z}$ and 
$\bar{\mathbf{Z}}$ should be interpreted as ``open'' covariant derivatives that operate on everything to their right.
For our purposes, only the $\mathbf{U}$ matrix will be present.

By using the decomposition \eqref{K-X} into \eqref{del L d phi2},
and expanding the log, we obtain
\eqali{
\label{log.power}
\frac{i}{2}\STr\log\left(\mathbf{K}-\mathbf{X}\right)\Big|_{\text{hard}}
&=\frac{i}{2}\STr\log\mathbf{K}\Big|_{\text{hard}}-\frac{i}{2}\sum^\infty_{n=1}\frac{1}{n}\STr\left[\left(\mathbf{K}^{-
1}\mathbf{X}\right)^{n}\right]\Big|_{\text{hard}}.
}
The contributions above can be separated in two:
the \emph{log-type} and the \emph{power-type} contributions, respectively.
The log-type contributions depend only on the heavy field propagators arising from 
the Lagrangian kinetic terms in the UV. The calculations of log-type terms yield contributions that 
depend only on the gauge fields associated with the heavy field. 
As such, this contribution vanishes when the heavy field is a gauge singlet. 
On the other hand, the power-type contributions depend on the other interaction terms of the UV Lagrangian.
In actual calculations, the functional part of the supertrace can be made explicit as\,\cite{HLM:1}
\eqali{
\label{STr.functional}
\left.\mathrm{STr}\,\mathcal{O}\left[P_{\mu}, U\right]\right|_{\text {hard }}= \pm\left.\int d^{d} x \int 
\frac{d^{d} q}{(2 \pi)^{d}} \operatorname{tr} \mathcal{O}\left[P_{\mu}-q_{\mu}, U\right]\right|_{\text {hard }}
\,,
}
where $\pm$ denotes the case of bosons and fermions respectively and the remaining trace $\tr[~~]$ is over the rest of 
the internal degrees of freedom such as gauge and Dirac space.

For multi-field theories, the whole $\mathbf{U}$ matrix contains various entries which can be organized schematically 
by opening the light fields $\phi$ into
\eq{
\varphi=\mtrx{\varphi_s\cr \varphi_f\cr \varphi_V}
\,,\quad
\bar{\varphi}=\mtrx{\bar{\varphi}_s& \bar{\varphi}_f& \varphi_V}\,,
}
where $s,f,V$ denotes scalars, fermions and vectors, respectively.
For complex scalars, we can further divide
\eq{
\varphi_s=\mtrx{s_i \cr s_i^*}\,,
\quad
\bar{\varphi}_s=\mtrx{s_i^\dag & s_i^\tp}\,,
}
and, analogously for fermions, 
\eq{
\varphi_f=\mtrx{f_i \cr f_i^c}\,,
\quad
\bar{\varphi}_s=\mtrx{\bar{f}_i & \overline{f^c_i}}\,.
}
The $(~)^\tp$ denotes the transpose in other spaces such as gauge.
When the scalars are real, the entry $s_i^*$ in $\varphi_s$ and $s_i^\tp$ in $\bar{\varphi}_s$ should be removed.
For fermions, four component spinors are used and for chiral fields in the SM, it is understood that dummy auxiliary 
chiral fields should be paired to form the four component spinors such as pairing a singlet $d_L'$ with the SM $d_R$ 
down quark singlet Weyl field\,\cite{cohen.zhang:20}.

\section{Tree level matching at $\Lcp$}
\label{sec:tree:Lcp}

To obtain the tree level EFT Lagrangian after integrating out the CP breaking scalars $s_i$ at $\Lcp$, 
we need the EOM for the heavy fields $s_i$ determined from the UV Lagrangian \eqref{lag:UV:mass}:
\eqali{
\label{EOM:si}
0=-\frac{\delta S}{\delta s_i}&= \Box s_i+M^2_{ij}s_j+H^\dag H\gamma_{ij}s_j+\frac{1}{2}\lambda'_{ijk} 
s_js_k+\frac{1}{3!}\lambda_{ijkl}s_js_ks_l
\cr&\quad
+\ 
H^\dagger H\gamma^\prime_i
+\Big(\bar{B}_LF_id_R+\bar{B}_LG_iB_R+h.c.\Big)\,.
}
We use the VLQ mass basis but at tree level the CP basis can be equally used.

The classical solution for $s_i$ can be expanded\,\footnote{%
For heavy scalar fields with interactions that are cubic or higher degree,  
the method of expanding the inverse of the operator accompanying the bilinear in the heavy fields\,\cite{HLM:1}
needs to be applied carefully.
}
\eq{
\label{si:classical}
(s_i)_c=s_i^{(2)}+s_i^{(3)}+s_i^{(4)}+\cdots\,,
}
where $s_i^{(n)}$ are operators of dimension $n$.
Collecting terms of operator dimension from two up to six, we obtain
\eqali{
\label{EOM:si:order}
0&=M^2_{ij}s_j^{(2)}+\gamma'_iH^\dag H\,,
\cr
0&=M^2_{ij}s_j^{(3)}+\big(\bar{B}_LF_id_R+\bar{B}_LG_iB_R+h.c.\big)
\,,
\cr
0&=\Box s_i^{(2)}+M^2_{ij}s_j^{(4)}+H^\dag H\gamma_{ij}s_j^{(2)}+\ums{2}\lambda'_{ijk}s_j^{(2)}s_k^{(2)}
 \,,
 \cr
 0&=\Box s_i^{(3)}+M^2_{ij}s_j^{(5)}+H^\dag H\gamma_{ij}s_j^{(3)}+\lambda'_{ijk}s_j^{(2)}s_k^{(3)}
 \,,
 \cr
 0&=\Box s_i^{(4)}+M^2_{ij}s_j^{(6)}+H^\dag 
H\gamma_{ij}s_j^{(4)}+\lambda'_{ijk}\Big(s_j^{(2)}s_k^{(4)}+\ums{2}s_j^{(3)}s_k^{(3)}\Big)
+\ums{6}\lambda_{ijkl}s_j^{(2)}s_k^{(2)}s_l^{(2)}
\,.
}
The solution of \eqref{EOM:si:order} order by order is 
\eqali{
\label{si:classical:order}
s_j^{(2)}&=C^{sH}_jH^\dag H\,,
\cr
s_j^{(3)}&=
\bar{B}_L(C^{sBd}_jd_R+C^{sBB}_jB_R)+h.c.
\,,
\cr
s_j^{(4)}&=C^{s\Box H}_j\Box(H^\dag H)
+
C^{sH^4}_j(H^\dag H)^2
\,,
\cr
 s_j^{(5)}&=\Box(\bar{B}_LC^{s\Box Bd}_jd_R+\bar{B}_LC^{s\Box BB}_jB_R)
 +\bar{B}_L(C^{sBdH}d_R+ C^{sBBH}B_R)H^\dag H
 +h.c.
 \,,
\cr
s_j^{(6)}&=-M^{-2}_{ji}\Box s_i^{(4)}
-M^{-2}_{ji}\Big(\gamma_{ik}-\lambda'_{ikl}(M^{-2}\gamma')_l\Big)H^\dag H s_k^{(4)}
\cr&\quad
 -\ums{2}M^{-2}_{ji}\lambda'_{ikl}s_k^{(3)}s_l^{(3)}
 -\ums{6}M^{-2}_{ji}\lambda_{iklm}s_k^{(2)}s_l^{(2)}s_m^{(2)}
 \,,
}
with coefficients
\eqali{
C^{sH}_j&\equiv -M^{-2}_{ji}\gamma'_i
\,,
\cr
[C^{sBd}_j]^{ap}&\equiv -M^{-2}_{ji}[F_i]^{ap}
\,,
\cr
[C^{sBB}_j]^{ap}&\equiv -M^{-2}_{ji}[G_i]^{ap}
\,,
\cr
C^{s\Box H}_{j}&\equiv +M^{-4}_{ji}\gamma'_i
\,,
\cr
C^{sH^4}_j&\equiv 
+M^{-2}_{ji}\gamma_{ik}
M^{-2}_{kl}\gamma'_l
-\ums{2}M^{-2}_{ji}\lambda'_{ikl}M^{-2}_{kk'}\gamma'_{k'}M^{-2}_{ll'}\gamma'_{l'}
\,,
\cr
[C^{s\Box Bd}_j]^{ap}&\equiv +M^{-4}_{ji}[F_i]^{ap}
\,,
\cr
[C^{s\Box BB}_j]^{ap}&\equiv +M^{-4}_{ji}[G_i]^{ap}
\,,
\cr
[C^{sBdH}_j]^{ap}&=+M^{-2}_{ji}\left(\gamma_{ik}-\lambda'_{ilk}(M^{-2}\gamma')_l\right)M^{-2}_{km}[F_m]^{ap}\,,
\cr
[C^{sBBH}_j]^{ap}&=+M^{-2}_{ji}\left(\gamma_{ik}-\lambda'_{ilk}(M^{-2}\gamma')_l\right)M^{-2}_{km}[G_m]^{ap}\,,\cr
}
We use $M^{-2}\equiv (M^2)^{-1}$ and $M^{-4}\equiv (M^2M^2)^{-1}$.
Only the solutions up to $s_j^{(4)}$ are needed for tree-level matching.
The higher order $s_j^{(5)},s_j^{(6)}$ are needed for one-loop matching if we want dimension six operators.

The EFT Lagrangian matched at tree level is
\eqali{
\label{lag:eft:mass}
\lag_{\rm EFT}^{(0)}&=\text{(kinetic)}
    -\Big(\bar{q}_{L}Y^u\tilde{H}u_{R}+\bar{q}_LY^d Hd_{R}
    +\bar{q}_LY^B HB_{R}+h.c.\Big)
\cr
&\quad 
-\Big(\bar{B}_{L}M^BB_{R}
    +h.c.\Big) -V(|H|^2)
    +\sum_{k}C_k\cO_k
\,.
}
We list in table \ref{tab:op:mass:tree} the generated operators $\cO_k$ up to dimension 6 and the respective Wilson 
coefficients.
For completeness, we also list in table \ref{tab:op:cp:tree} the generated operators in the CP basis \eqref{lag:UV:cp}.
We use the same symbols for the quark fields but they differ by the transformation \eqref{WR:w} between the two bases.
Note that for Nelson-Barr theories various Wilson coefficients are related by the relation \eqref{Gi:Fi.YB}.
For example, the seventh and eighth Wilson coefficients in table \ref{tab:op:mass:tree} are related by
\eq{
\label{BBHH->BdHH}
C_{BBHH}=C_{BdHH}{Y^d}^{-1}Y^B\,.
}
This is also reflected in the fewer number of coefficents in table \ref{tab:op:cp:tree}.
\begin{table}[ht]
\[
\begin{array}{|c|c|c|c|c|}
\hline
 & \text{Operator} &  \text{Wilson coefficient}&\text{Op.\ dim.}\\
 \hline
\rule{0ex}{3ex}
\mathcal{O}^{ara'r'}_{BdBd} & \bar{B}_L^a d_R^r\bar{B}_L^{a'}d_R^{r'}  &  
    \frac{1}{2}M^{-2}_{ik} [F_{i}]^{ar}[F_{k}]^{a'r'}   &6
\\[.5ex]
\mathcal{O}^{raa'b'}_{dBBd} 
&\bar{d}^r_RB^a_L\bar{B}^{a'}_Ld^{r'}_R   
& M^{-2}_{ik}[F_{i}^\dag]^{ra}[F_{k}]^{a'r'} &6

\\[.5ex]
\mathcal{O}^{aba'b'}_{B_LB_RB_LB_R} & \bar{B}_L^a B_R^b\bar{B}_L^{a'}B_R^{b'}  &  
    \frac{1}{2}M^{-2}_{ik} [G_{i}]^{ab}[G_{k}]^{a'b'}  &6 
\\[.5ex]
\mathcal{O}^{aba'b'}_{B_RB_LB_LB_R} & \bar{B}_R^a B_L^b\bar{B}_L^{a'}B_R^{b'}  &  
    M^{-2}_{ik} [G^\dag_{i}]^{ab}[G_{k}]^{a'b'}  &6
\\[.5ex]
\mathcal{O}^{raa'b'}_{dBBB} 
&\bar{d}^r_RB^a_L\bar{B}^{a'}_LB^{b'}_R   
& \frac{1}{2}M^{-2}_{ik}[F_{i}^\dag]^{ra}[G_{k}]^{a'b'} &6 
\\[.5ex]
\mathcal{O}^{bra'b'}_{BBBd} 
&\bar{B}^b_Ld^r_R\bar{B}^{a'}_LB^{b'}_R   
& \frac{1}{2}M^{-2}_{ik}[F_{i}]^{br}[G_{k}]^{a'b'} &6

\\[.5ex]
\mathcal{O}^{ar}_{BdHH} &\bar{B}^a_L d^r_R(H^\dagger H)  &   M^{-2}_{ik}[F_{i}]^{ar}\gamma'_k&5

\\[.5ex]
\mathcal{O}^{ab}_{BBHH} &\bar{B}^a_L B^b_R(H^\dagger H)  &   M^{-2}_{ik}[G_{i}]^{ab}\gamma'_k&5

\\[.5ex]
\mathcal{O}_{H\Box } & (H^\dagger H)\Box(H^\dagger H) 
    &-\frac{1}{2}\gamma'_i M^{-4}_{ik}\gamma'_k  &6

\\[.5ex]
\cO_{H} & (H^\dag H)^3 
    & -\frac{1}{2}\gamma^\prime_i M^{-2}_{ij}\gamma_{jl}M^{-2}_{lk}\gamma^\prime_k
    +\ums{6}\lambda'_{ijk}M^{-2}_{il}\gamma^\prime_l M^{-2}_{jm}\gamma^\prime_m M^{-2}_{kn}\gamma^\prime_n&6

    \\[.5ex]
 \mathcal{O}_{\delta \lambda_H} & (H^\dagger H)^2  
    &\frac{1}{2}\gamma^\prime_i M^{-2}_{ik}\gamma^\prime_k&4
\\[.7ex]
\hline
\end{array}
\]
\caption{\label{tab:op:mass:tree}%
Operators and Wilson coefficients for the UV Lagrangian \eqref{lag:UV:mass} in VLQ mass basis and the general potential 
in \eqref{VS:general} and \eqref{VHS:general} after integrating out $s_i$ at tree level.
For non-hermitian operators, the presence of their hermitean conjugates is understood.
The operators follow the convention of Ref.\,\cite{Jenkins:2017jig}.
}
\end{table}
\begin{table}[ht]
\[
\begin{array}{|c|c|c|c|c|}
\hline
 & \text{Operator} &  \text{Wilson coefficient}&\text{Op.\ dim.}\\
\hline
\rule{0ex}{3ex}
\mathcal{O}^{ara'r'}_{BdBd} & \bar{B}_L^a d_R^r\bar{B}_L^{a'}d_R^{r'}  &  
    \frac{1}{2}M^{-2}_{ik} [\cF_i]^{ar}[\cF_k]^{a'r'}  & 6
\\[.5ex]
\mathcal{O}^{raa'r'}_{dBBd} 
&\bar{d}^r_RB^a_L\bar{B}^{a'}_Ld^{r'}_R   
& M^{-2}_{ik}[\cF_i^\dag]^{ra}[\cF_k]^{a'r'} &6
\\[.5ex]
\mathcal{O}^{ar}_{BdHH} &\bar{B}^a_L d^r_R(H^\dagger H)  &   M^{-2}_{ik}[\cF_i]^{ar}\gamma'_k&5
\\[.5ex]
\mathcal{O}_{H\Box } & (H^\dagger H)\Box(H^\dagger H) 
    &-\frac{1}{2}\gamma'_i M^{-4}_{ik}\gamma'_k  &6
\\[.5ex]
\cO_{H} & (H^\dag H)^3 
    & -\frac{1}{2}\gamma^\prime_i M^{-2}_{ij}\gamma_{jl}M^{-2}_{lk}\gamma^\prime_k
    +\ums{6}\lambda'_{ijk}M^{-2}_{il}\gamma^\prime_l M^{-2}_{jm}\gamma^\prime_m M^{-2}_{kn}\gamma^\prime_n&6
    \\[.5ex]
 \mathcal{O}_{\delta \lambda_H} & (H^\dagger H)^2  
    &\frac{1}{2}\gamma^\prime_i M^{-2}_{ik}\gamma^\prime_k&4
\\[.7ex]
\hline
\end{array}
\]
\caption{\label{tab:op:cp:tree}%
Operators and Wilson coefficients for the UV Lagrangian \eqref{lag:UV:cp} in CP basis and the general potential in 
\eqref{VS:general} and \eqref{VHS:general} after integrating out $s_i$ at tree level.
For non-hermitian operators, the presence of their hermitean conjugates is understood.
The operators follow the convention of Ref.\,\cite{Jenkins:2017jig}.
}
\end{table}

For Higgs-only operators, the contributions are analogous to the addition of one singlet scalar to the SM.
In that case, our results match e.g. Ref.\,\cite{cohen.zhang:20}.

\section{One-loop matching at $\Lcp$}
\label{sec:1L:Lcp}

In the Lagrangian \eqref{lag:UV:mass}, the relevant interactions beyond the SM are the ones with couplings $F_i,G_i$ 
for 
the Yukawas and $V_{HS},V_S$ in the scalar potential.
Following Ref.\,\cite{cohen.zhang:20}, we can write the relevant $U$ matrices:
\eqali{
\label{U:1}
U^{[2]}_{Bd}&=\mtrx{F_is_iR & 0 \cr 0 & F_i^*s_iL}\,,
\hs{3.5em}
U^{[2]}_{BB}=\mtrx{G_is_iR +G_i^\dag s_iL& 0 \cr 0 & G_i^*s_iL+G_i^\tp s_iR}\,,
\cr
U^{[1]}_{qB}&=\mtrx{Y^BRH & 0 \cr 0 & {Y^{B}}^*LH^*}\,,
\hs{2em}
U^{[\sfrac{3}{2}]}_{s_id}=\mtrx{\overline{B}F_iR & \overline{B^c}F_i^*L}\,,
\cr
U^{[\sfrac{3}{2}]}_{s_iB}&=\mtrx{\overline{B}(G_iR+G_i^\dag L)+\bar{d}F_i^\dag L, & \overline{B^c}(G_i^\tp 
R+G_i^*L)+\overline{d^c}F_i^\tp R}\,,
}
where we use the chiral projectors $R=\ums{2}(\id+\gamma_5)$, $L=\ums{2}(\id-\gamma_5)$
and recall that the heavy fields $s_i$ should be replaced by the classical solution in \eqref{si:classical}.
The numbers in brackets denote the minimal operator dimension and only the ones involving the classical solution for 
$s_i=(s_i)_c$ have contributions of higher dimensions.
Conservation of fermion number and the sole appearance of fermions in bilinears (e.g. no quartics) lead to 
$U_{f'f}=\overline{U_{ff'}}$ where $\overline{A}=\gamma_0A^\dag\gamma_0$ with the dagger acting on all Dirac, gauge and 
family spaces.
For scalar $s_i$ and fermion $f$ we have similarly
$U_{sf}=\overline{U_{fs}}\equiv U_{fs}^\dag\gamma_0$.
Note that $\overline{R}=L$ and chirality is exchanged.
The rest of $U$ matrices are analogous to the SM augmented by a singlet scalar which is calculated in 
Ref.\,\cite{cohen.zhang:20}.
For example,
\eqali{
\label{U:2}
U^{[\sfrac{3}{2}]}_{HB}&=
\mtrx{0 & \bar{q}^c{Y^{B}}^*L
    \cr \bar{q}Y^BR & 0}\,,
}
where the $SU(2)_L$ indices of the doublet $\bar{q}$ are understood as a column vector.

We can obtain the scalar-scalar $U$ matrices from $V_S$ in \eqref{VS:general} and $V_{HS}$ 
in \eqref{VHS:general}, which yields
\eqali{
\label{U:3}
U^{[2]}_{s_is_j}&=(H^\dag H)\gamma_{ij}+\lambda'_{ijk}s_k+\ums{2}\lambda_{ijkl}s_ks_l\,,
\cr
U_{s_iH}^{[1]}&=
\mtrx{
(\gamma_{ij}s_j+\gamma'_i)H^\dag
&
(\gamma_{ij}s_j+\gamma'_i)H^\tp
}
\,.
}
$U_{s_js_i}$ is symmetric and $U_{H s_i}=(U_{s_iH})^\dag$.
Similarly, the Higgs-Higgs $U$ matrix is
\eqali{
U^{[2]}_{H H}=&\Big[\gamma_i's_i+\ums{2}\gamma_{ij}s_is_j\Big]\left(\begin{array}{ll}
\mathbb{1} & 0 \\
0 & \mathbb{1}
\end{array}\right)
+(\text{quartic})
\,,
}
where quartic means the contribution from the Higgs quartic coupling $|H|^4$ which can be found in 
Ref.\,\cite{cohen.zhang:20}.

\subsection{Threshold corrections}

The corrections to the renormalizable operators in the Green basis\footnote{%
The Green basis refer to the natural basis of operators that arises when we integrate heavy fields.
}
are 
\eqali{
\label{lag:1l:dim4}
\lag_{\rm EFT}^{\text{1-$\ell$}} & \supset\delta Z_H(D_\mu H)^\dagger(D^\mu H)+\delta\mu_H H^\dagger H+\delta 
\lambda(H^\dagger H)^2
\\
&+\sum_{f=d_R,B_L,B_R}\bar{f}\delta Z_fi\slashed{D}f+\bar{B}_R\delta Z_{Bd}i\slashed{D}d_R+\bar{d}_R\delta 
Z^\dagger_{Bd}i\slashed{D}B_R
\\
&+\Big(\bar{B}_L\delta M^B_GB_R+ \bar{B}_L C^{Bd}d_R+h.c.\Big)
\\
&+\Big(\bar{q}_{L}\delta Y^d_GHd_{R}+\bar{q}_{L}\delta Y^u_G\tilde{H}u_{R}+\bar{q}_{L}\delta Y^B_GHB_{R}
+h.c.\Big)\,.
}
The label $G$ refers to the Green basis.

We will make use of\,\cite{Zhang:2021jdf}
\begin{equation}
\label{def:log}
L_i\equiv \log\left(\frac{\mu^2}{M^2_i}\right)\,,\quad
L_{ij}\equiv L_i\delta_{ij}\,,
\end{equation}
where $\mu=\Lcp$ is the matching scale at the CP breaking scale. Typically, $\Lcp\sim u_i\sim M_i$.
In this notation, the generalization to bases where the scalar mass matrix $M^2$ is nondiagonal is straightforward and 
in such a case $L_{ij}$ becomes a nondiagonal matrix accordingly.

We start with the corrections to the Higgs sector:
\subeqali[]{
16\pi^2\delta Z_H&=\ums{2}\gamma_i'M^{-2}_{ij}\gamma_j'\,,
\\
16\pi^2\delta\mu^2_H&=
    \ums{2}M^2_{ij}(\delta_{jk}+L_{jk})\big[\gamma_{ki}-\lambda_{kil}(M^{-2}\gamma')_l\big]
    +(\delta_{ij}+L_{ij})\gamma_i'\gamma_j' 
\,,
}
where summation of repeated indices is implicit and $(M^2)_{ij}=M^2_i\delta_{ij}$ in the diagonal basis;
similarly $(M^{-2}\gamma')_i=M^{-2}_{ij}\gamma'_j$.
Notice the contribution of the heavy mass $M^2$ to $\delta\mu_H^2$ which is reminiscent of the hierarchy problem.
These expressions for $\delta Z_H$ and $\delta \mu_H$ match Ref.\,\cite{cohen.zhang:20} neglecting $\mu^2_H/M^2$.
We omit the expression for $\delta\lambda$ which involves many supertraces. For only one singlet scalar, it can be
found in Ref.\,\cite{cohen.zhang:20}.
The other wave-function corrections are
\subeqali[eff.deltaZ]{
\label{deltaZ:dR:eff}
16\pi^2[\delta Z_{d_R}]^{pr}&=\ums{2}[F_i^\dag(\ums{2}\delta_{ij}+L_{ij})F_j]^{pr}\,,
\\
\label{deltaZ:BR:eff}
16\pi^2[\delta Z_{B_R}]^{ab}&=\ums{2}[G_i^\dag(\ums{2}\delta_{ij}+L_{ij})G_i]^{ab}\,,
\\
\label{deltaZ:Bd:eff}
16\pi^2[\delta Z_{Bd}]^{ar}&=\ums{2}[G_i^\dag(\ums{2}\delta_{ij}+L_{ij})F_i]^{ar}\,,
\\
\label{deltaZ:BL:eff}
16\pi^2[\delta 
Z_{B_L}]^{ab}&=\ums{2}[F_i(\ums{2}\delta_{ij}+L_{ij})F_j^\dag+G_i(\ums{2}\delta_{ij}+L_{ij})G_j^\dag]^{ab}\,.
}
Here and in the following, we omit corrections suppressed by $M^B/M_i$.

The correction to the fermion mass terms are
\subeqali[]{
\label{deltaMB:G}
16\pi^2[\delta M^B_G]^{ab}&=[G_i 
{M^B}^\dag(\delta_{ij}+L_{ij})G_j-\ums{2}M^2_{jn}(\delta_{ni}+L_{ni})\lambda'_{ijk}M^{-2}_{kl}G_l]^{ab}
\,,
\\
16\pi^2[C^{Bd}]^{ar}&=[G_i 
{M^B}^\dag(\delta_{ij}+L_{ij})F_j-\ums{2}M^2_{jn}(\delta_{ni}+L_{ni})\lambda'_{ijk}M^{-2}_{kl}F_l]^{ar}
\,,
}
where $M^B=\hat{M}^B$ is diagonal.

The corrections to the Yukawa couplings are
\subeqali[]{
16\pi^2[\delta Y^d_G]^{pr}&=(M^{-2}\gamma')_j(\delta_{jk}+L_{jk})[Y^B {M^B}^\dag F_k]^{pr}\,,
\\
\label{deltaYB:G}
16\pi^2[\delta Y^B_G]^{pb}&=(M^{-2}\gamma')_j(\delta_{jk}+L_{jk})[Y^B {M^B}^\dag G_k]^{pb}\,.
}
We neglect corrections of order $\mu_H^2/M^2$ which would correspond here to adding $(\delta_{ij}+\mu_H^2M^{-2}_{ij})$ 
in 
the middle.

\subsection{Field redefinition}
\label{Lcp:field.redef}

Considering the one-loop corrections \eqref{lag:1l:dim4} to the tree level Lagrangian \eqref{lag:UV:mass}, we need to 
perform the following field redefinitions to transform the kinetic terms to canonical form and make the VLQs to have 
definite mass:
\eqali{
\label{field.redef:Z}
H &\rightarrow \left[1+\delta Z_{H} \right]^{-1/2}H\,,\\
B_L &\rightarrow \left[\id+\delta Z_{B_L} \right]^{-1/2}B_L\,,\\
\begin{pmatrix}
    B_R\\
    d_R
\end{pmatrix}&\rightarrow
\begin{pmatrix}
   \id-\frac{1}{2}\delta Z_{B_R}&-\frac{1}{2}\delta Z_{Bd}\\
    -\frac{1}{2}\delta Z^\dagger_{Bd} &\id_3-\frac{1}{2}\delta Z_{d_R}
\end{pmatrix}
\mtrx{\id &\zeta\\
    -\zeta^\dagger&\id}
\begin{pmatrix}
    B_R\\
    d_R
\end{pmatrix}\,,
}
where
\eq{
\zeta\equiv +{M^B}^{-1}C^{Bd}+\ums{2}\delta Z_{Bd}\,.
}
Then the EFT Lagrangian up to dimension four will be
\eqali{
\label{lag:1-L:redefined}
\lag_{\rm EFT}^{\text{1-$\ell$}}&\supset \text{(kinetic)}
-\Big(\bar{q}_{L}Y^u_{\rm eff}\tilde{H}u_{R}+\bar{q}_L Y^d_{\rm eff} Hd_{R}+\bar{q}_L Y^B_{\rm eff} HB_{R} 
+\bar{B}_{L} M^B_{\rm eff}B_{R}
+h.c.\Big)
\cr
&\quad 
-V(|H|^2)
\,,
}
with effective couplings
\subeqali[eff.corrected]{
[Y^u_{\rm eff}]^{pr}&=[Y^u]^{pr}\big(1-\ums{2}\delta Z_H\big)\,,
\\
\label{Yd:eff}
[Y^d_{\rm eff}]^{pr}&=[Y^d(\id_3-\ums{2}\delta Z_H\id_3-\ums{2}\delta Z_{d_R})-\delta Y^d_G+Y^B{M^B}^{-1}C^{Bd}]^{pr}\,,
\\
\label{YB:eff}
[Y^B_{\rm eff}]^{pb}&=[Y^B(\id_n-\ums{2}\delta Z_H\id_n-\ums{2}\delta Z_{B_R})-\delta Y^B_G-Y^d(\delta Z_{Bd}^\dag
+{C^{Bd}}^\dag {M^B}^{\dag -1})]^{pb}\,,
\\
\label{MB:eff}
[M^B_{\rm eff}]^{ab}&=[M^B-\delta M^B_G-\ums{2}\delta Z_{B_L}M^B-\ums{2}M^B\delta Z_{B_R}]^{ab}\,.
}
Due to the contribution of Fig.\,\ref{diag:deltaMB}(b) to \eqref{deltaMB:G}
in \eqref{MB:eff}, we see that the limit $M^B\to 0$ is technically natural only if $\lambda'_{ijk}$ or 
$G_l$ is equally suppressed.
This is natural in the Nelson-Barr setting because $\lambda'_{ijk}\sim \Lcp$ and $G_l\sim M^B/\Lcp$.
In terms of the original Lagrangian \eqref{lag:UV:cp}, this corresponds to the limit of a chiral symmetry for $B_L$
which forbids $\cM^{Bd},\cF_i,\cM^B$.

\subsection{Threshold correction to $\btheta$}
\label{sec:Lcp:theta-bar}

Within the SM, the $\theta$-term in QCD is
\eq{
\lag=\theta\frac{g_s^2}{64\pi^2}\varepsilon^{\mu\nu\alpha\beta}G^a_{\mu\nu}G^a_{\alpha\beta}\,,
}
with $\varepsilon^{0123}=1$, 
while the basis invariant phase is
\eq{
\btheta=\theta-\arg\det(Y^d)-\arg\det(Y^u)\,,
}
as chiral transformations induce a shift in $\theta$.
For example, with a chiral transformation on a single up-type righthanded quark,
\eq{
\label{rephasing.u:alpha}
u_{1R}\to e^{i\alpha}u_{1R}
\implies \theta\to \theta+\alpha\,,
}
while $\btheta$ is invariant by construction.

In Nelson-Barr theories, prior to spontaneous CP violation (SCPV), $\theta=0$ and there is no phase in any quark Yukawa 
coupling.
After SCPV, matching everything at tree level, the NB setting still ensures real determinant for the Yukawas $Y^d$ and 
$Y^u$; cf.\,\eqref{WR:YdYB}.
In general, things are different already at one-loop\,\cite{nelson,barr,BBP}.

\begin{figure}
\raisebox{7em}{{\small (a)}}
\hspace{-2em}
\includegraphics[scale=.75]{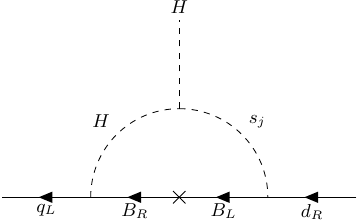}
\hspace{1em}
\raisebox{7em}{{\small (b)}}
\hspace{-2em}
\includegraphics[scale=.75]{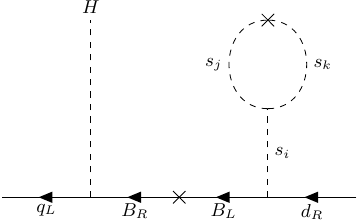}
\hspace{1em}
\raisebox{7em}{{\small (c)}}
\hspace{-2em}
\includegraphics[scale=.75]{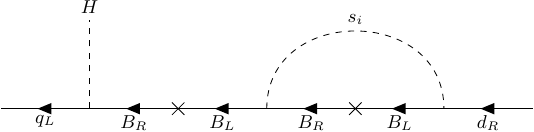}
\caption{\label{diag:deltaYd}%
One-loop threshold contributions to $Y^d_{\rm eff}$ in \eqref{Yd:eff} relevant to $\btheta$ when integrating out the 
scalars.
}
\end{figure}

\begin{figure}
\raisebox{5em}{{\small (a)}}
\hspace{-2em}
\includegraphics[scale=.75]{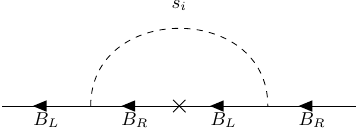}
\hspace{1em}
\raisebox{5em}{{\small (b)}}
\hspace{-2em}
\includegraphics[scale=.75]{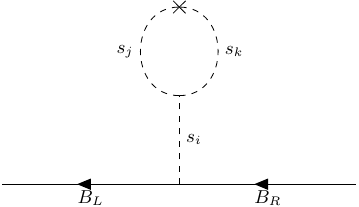}
\caption{\label{diag:deltaMB}%
One-loop threshold contributions to $M^B_{\rm eff}$ in \eqref{MB:eff} relevant to $\btheta$ when integrating out the 
scalars.
}
\end{figure}

Considering the two matching scales in \eqref{efts}, if we integrate out the NB-VLQs just at tree level,\footnote{%
We will correct this in Sec.\,\ref{sec:Mvlq}.}
the value of $\btheta$ after integrating out the CP breaking scalars at one-loop matched at $\Lcp$ is
\eq{
\label{theta:Lcp}
\btheta(\Lcp)=
\theta_d+\theta_B\,,
}
where
\eqali{
\label{thetad.thetaB}
\theta_d&=-\arg\det(Y^d_{\rm eff})\,,
\cr
\theta_B&=-\arg\det(M^B_{\rm eff})\,.
}
Note that the contribution from $Y^u$ is real.
The contribution to $\theta_B$ occurs because $\det M^B_{\rm eff}$ has a phase $e^{-i\theta_B}$ which can be rotated 
away, e.g., by the chiral transformation
\eq{
\label{rephasing:BR}
B_{aR}\to e^{i\theta_B/n}B_{aR}\,,\quad a=1,...,n.
}
Analogously to \eqref{rephasing.u:alpha}, this chiral global rephasing induces $\theta=\theta_B$ while $Y^d_{\rm eff}$ 
will correspond to the SM down-type Yukawa coupling after integrating out the NB-VLQs at tree level which is best 
performed in the basis where $M^B_{\rm eff}e^{i\theta_B/n}$ is diagonal. But this diagonalization within $SU(n)$ does 
not modify $\theta$ any further.
The coupling $Y^B_{\rm eff}$ in \eqref{lag:1-L:redefined}, and possibly some Wilson coefficients of higher dimensional 
operators, will be also modified by the rephasing \eqref{rephasing:BR}.
We show respectively in Fig.\,\ref{diag:deltaYd} and Fig.\,\ref{diag:deltaMB} the relevant contributions from $Y^d_{\rm 
eff}$ and $M^B_{\rm eff}$ to $\theta_d$ and $\theta_B$ in \eqref{thetad.thetaB}.

Now let us calculate the phases in \eqref{thetad.thetaB}.
These phases would vanish if only tree level matching were considered at $\Lcp$.
Considering the one-loop matched couplings in \eqref{eff.corrected}, with hermitean wave function corrections in 
\eqref{eff.deltaZ}, the only contributions to $\theta_d$ and $\theta_B$ are
\eqali{
\label{theta.d.B}
\theta_d&=\im\tr[{Y^d}^{-1}(\delta Y^d_G-Y^B{M^B}^{-1}C^{Bd})]\,,
\cr
\theta_B&=\im\tr[{M^B}^{-1}\delta M^B_G]\,.
}

For a Nelson-Barr theory, several terms cancel due to the relation \eqref{Gi:Fi.YB} and
the contribution that remains is due to the diagram in Fig.\,\ref{diag:deltaYd}a which simply gives
\eq{
\label{thetabar:int.si}
16\pi^2
\btheta(\Lcp)=
(M^{-2}\gamma')_i(\delta_{ij}+L_{ij})
\im\tr[{Y^d}^{-1}Y^B{M^B}^\dag F_j]
\,.
}
After rewriting it using the quantities in the CP basis, the end result is the simple formula
\eq{
\label{thetabar:int.si:cp}
16\pi^2\btheta(\Lcp)=
(M^{-2}\gamma')_i(\delta_{ij}+L_{ij})\im\tr[{\cM^{Bd}}^\dag \cF_j]\,.
}
For the leading correction in $\mu^2_H/M^2_i$, we need to add $(\delta_{ki}+\mu^2_H M^{-2}_{ki})$ in the middle.
As $M^2\sim \Lcp^2$ and $\gamma'_i\sim \Lcp$, this contribution is suppressed as $\Lcp^{-1}$.
Also note that $\cF_i(M^{-2}\gamma')_i$ is the Wilson coefficient of $\bar{B}_Ld_R|H|^2$ from tree level matching in 
the 
CP basis; cf.\ Table~\ref{tab:op:cp:tree}.

\subsection{Dimension five}
\label{sec:dim.5}

The dimension five operators that are directly generated at one-loop are 
\eqali{
\label{EFT:s:dim5}
\lag_{\rm EFT}^{\text{1-$\ell$}}
&\supset
+\bar{q}_Li\cev{\slashed{D}}H C_{DqHB} B_L
+\bar{q}_L H C_{qHDB} i\slashed{D}B_L
\\
&\quad
+\bar{B}_L\cev{\slashed{D}}C_{DBDB}\slashed{D}B_R
+\bar{B}_L\cev{\slashed{D}}C_{DBDd}\slashed{D}d_R
\\
&\quad
+\bar{B}_LC^\sigma_{BB} \sigma{\cdot}F B_R
+\bar{B}_LC^\sigma_{Bd} \sigma{\cdot}F d_R
+h.c.\,,
}
where $\sigma{\cdot}F\equiv\sigma^{\mu\nu}F_{\mu\nu}$ with
\eq{
\label{F.sigma}
F_{\mu\nu}\equiv -i[D_\mu,D_\nu]
=g'\big({-}\ums{3}\big)B_{\mu\nu}+g_s G_{\mu\nu}\,.
}
The factor $-1/3$ is the hypercharge of $d_R$ or $B_R$.
We omit the one-loop corrections to the operators $\bar{B}_Ld_R|H|^2$ and $\bar{B}_LB_R|H|^2$ which are already 
generated at tree level; see Table\,\ref{tab:op:mass:tree}.
We list below the coefficients that are generated only at one-loop.
\subeqali[]{
16\pi^2[C_{DqHB}]^{pb}&=
-\ums{2}(M^{-2}\gamma')_k
\big(\ums[3]{2}\delta_{kj}+L_{kj}\big)
[Y^dF_j^\dag+Y^BG_j^\dag]^{pb}\,,
\\
16\pi^2[C_{qHDB}]^{pb}&=-\ums{2}(M^{-2}\gamma')_i[Y^dF_i^\dag+Y^BG_i^\dag]^{pb}\,.
}
\subeqali[]{
16\pi^2[C_{DBDB}]^{ab}&=\ums{2}M^{-2}_{ij}[G_i{M^B}^\dag G_j]^{ab}\,,
\\
16\pi^2[C_{DBDd}]^{ar}&=\ums{2}M^{-2}_{ij}[G_i{M^B}^\dag F_j]^{ar}\,.
}
\subeqali[dipole.1]{
16\pi^2[C^\sigma_{BB}]^{ab}&=\ums{2}M^{-2}_{ik}(\ums[3]{2}\delta_{kj}+L_{kj})[G_i {M^B}^\dag G_j]^{ab}\,,
\\
16\pi^2[C^\sigma_{Bd}]^{ar}&=\ums{2}M^{-2}_{ik}(\ums[3]{2}\delta_{kj}+L_{kj})[G_i {M^B}^\dag F_j]^{ar}\,.
}
We list in appendix \ref{ap:supertraces} the supertraces that lead to these operators.

We should note that $\slashed{D}B_L$ or $\slashed{D}B_R$ can be removed by the equations of motion for $B_{L,R}$:
\eqali{
\label{eom:B}
i\slashed{D}B_R&={M^B}^\dag B_L+H^\dag {Y^B}^\dag q_L\,,
\cr
i\slashed{D}B_L&={M^B} B_R\,,
}
which allow the reduction of the operator dimension by one unit.
In contrast, the equations of motion for the other SM fermions do not change the operator dimension.
The use of EOMs is justified through field redefinitions and the equivalence theorem\,\cite{Criado:2018sdb}.
The elimination of an operator of order $\Lcp^{-k}$ through the EOMs only leads to the correct equivalent effective 
Lagrangian at the same order and higher order modifications need to be tracked explicitly with the field redefinition.
We show some of the necessary redefinitions in appendix \ref{ap:field.redef} and also show two contributions 
appearing at order $(16\pi^2)^{-1}\Lcp^{-2}$ that cannot be tracked by the use of EOMs.
These need to be taken into account for a full matching at order $(16\pi^2)^{-1}\Lcp^{-2}$.

Applying \eqref{eom:B} into \eqref{EFT:s:dim5}, we obtain
\eqali{
\label{EFT:s:dim5:eom}
\lag_{\rm EFT}^{\text{1-$\ell$}}
&\supset
+\bar{q}_Li\cev{\slashed{D}}H C_{DqHB} B_L
+\bar{q}_LH\big(C_{qHDB}M^B+Y^BC^\dag_{DBDB}M^B\big)B_R
\\
&\quad
+\bar{B}_LC^\sigma_{BB} \sigma{\cdot}F B_R
+\bar{B}_LC^\sigma_{Bd} \sigma{\cdot}F d_R
+h.c.\,,
}
where we neglected $(M^B)^2/M^2$ corrections to $\delta Z_{Bd}$ in \eqref{deltaZ:Bd:eff} and $\delta M^B_G$ in 
\eqref{deltaMB:G}.
The second term in \eqref{EFT:s:dim5:eom} leads to corrections to $\delta Y^B_G$ in \eqref{deltaYB:G}.
The corrected coupling becomes
\eqali{
\label{deltaYB:G'}
16\pi^2[\delta Y^{\prime B}_{G}]^{pb}&=(M^{-2}\gamma')_i(\delta_{jk}+L_{jk})[Y^B {M^B}^\dag G_k]^{pb}
\cr
&\quad
-\ums{2}(M^{-2}\gamma')_i[(Y^dF_i^\dag+Y^BG_i^\dag)M^B]^{pb}
\,,
}
where we only keep the correction coming from $C_{qHDB}$ which is of the same order; 
the correction coming from $C_{DBDB}$ is further suppressed by another factor of $M^B/M_i$.
This contribution does not change our calculation for $\delta\btheta$ in \eqref{}.
By also using the equation of motion for $\slashed{D}q_L$ in the first term of \eqref{EFT:s:dim5:eom}, we generate the 
hermitean conjugate of the dimension five operators $\bar{B}_LB_R|H|^2$ and $\bar{B}_Ld_R|H|^2$ which are already 
generated at tree-level.
So the dipole operators involving $\sigma{\cdot}F$ in \eqref{EFT:s:dim5:eom} are the only new dimension five operators 
generated at one-loop:
\eqali{
\label{EFT:dim5:dipole}
\lag_{\rm EFT}^{\text{1-$\ell$}}\big|_{\rm new}
&=
+\bar{B}_LC^{\sigma\rm eff}_{BB} \sigma{\cdot}F B_R
+\bar{B}_LC^{\sigma\rm eff}_{Bd} \sigma{\cdot}F d_R
+h.c.\,.
}

The full contribution to the dipole operators, however, require the calculation of dimension 6 operators because some 
of 
them may turn into dimension five with the use of \eqref{eom:B}.
The relevant dimension 6 operators are\footnote{
We use a basis similar to Refs.\cite{Jenkins:2017dyc,bilenky}.
}
\eqali{
\label{EFT:dim6:green}
\lag_{\rm EFT}^{\text{1-$\ell$}}
&\supset
\bar{B}(i\slashed{D})^3C_{BD^3B}B
+\Big\{\bar{B}_R(i\slashed{D})^3C_{BD^3d}d_R+h.c.\Big\}
\cr
&\quad
+\bar{B}(-i\cev{\slashed{D}}\sigma{\cdot}F+\sigma{\cdot}F i\slashed{D}) C^{\sigma D}_{BB}B
+\Big\{\bar{B}_R(-i\cev{\slashed{D}}\sigma{\cdot}F+\sigma{\cdot}F i\slashed{D}) C^{\sigma D}_{Bd}d_R
+h.c.\Big\}
\cr
&\quad
+\bar{B}\gamma^\beta(D^\alpha F_{\alpha\beta}) C^{DF}_{BB}B
+\Big\{
\bar{B}_R\gamma^\beta(D^\alpha F_{\alpha\beta}) C^{DF}_{Bd}d_R
+h.c.\Big\}\,,
\cr
}
with coefficients
\subeqali[]{
16\pi^2[C_{BD^3B}]^{ab}&=\ums{6}M^{-2}_{ij}[G_i^\dag G_jR+(G_iG_j^\dag+F_iF_j^\dag)L]^{ab}\,,
\\
16\pi^2[C_{BD^3d}]^{ar}&=\ums{6}M^{-2}_{ij}[G_i^\dag F_j]^{ar}\,,
\\
16\pi^2[C^{\sigma D}_{BB}]^{ab}&=-\ums{12}M^{-2}_{ij}[G_i^\dag G_jR+(G_iG_j^\dag+F_iF_j^\dag)L]^{ab}\,,
\\
16\pi^2[C^{\sigma D}_{Bd}]^{ar}&=-\ums{12}M^{-2}_{ij}[G_i^\dag F_j]^{ar}\,,
\\
16\pi^2[C^{DF}_{BB}]^{ab}&=-\ums{3}M^{-2}_{ik}\big(\ums[4]{3}\delta_{kj}+L_{kj}\big)[G_i^\dag 
G_jR+(G_iG_j^\dag+F_iF_j^\dag)L]^{ab}\,,
\\
16\pi^2[C^{DF}_{Bd}]^{ar}&=-\ums{3}M^{-2}_{ik}\big(\ums[4]{3}\delta_{kj}+L_{kj}\big)[G_i^\dag F_j]^{ar}\,.
}
Note that $C_{BD^3B},C^{\sigma D}_{BB},C^{\gamma DF}_{BB}$ contain the chiral projectors $R,L$.
After applying the equations of motion \eqref{eom:B}, the operators with coefficients $C^{\sigma D}_{BB}$ and 
$C^{\sigma 
D}_{Bd}$ will contribute to the dipole operators with coefficients \eqref{dipole.1} 
effectively changing them to
\subeqali[dipole.eff]{
16\pi^2[C^{\sigma \rm eff}_{BB}]^{ab}&=16\pi^2[C^{\sigma}_{BB}]^{ab}
-\ums{12}M^{-2}_{ij}[(G_iG_j^\dag+F_iF_j^\dag)M^B+M^BG_i^\dag G_j]^{ab}
\cr
&=
\ums{2}M^{-2}_{ik}\Big\{(\ums[3]{2}\delta_{kj}+L_{kj})[G_i{M^B}^\dag G_j]^{ab}
-\ums{6}[(G_iG_k^\dag+F_iF_k^\dag)M^B+M^BG_i^\dag G_k]^{ab}
\Big\}
\,,
\\
16\pi^2[C^{\sigma \rm eff}_{Bd} ]^{ar}&=16\pi^2[C^{\sigma}_{Bd} ]^{ar}
-\ums{12}M^{-2}_{ij}[M^BG_i^\dag F_j]^{ar}
\cr
&=
\ums{2}M^{-2}_{ik}\Big\{(\ums[3]{2}\delta_{kj}+L_{kj})[G_i{M^B}^\dag F_j]^{ar}
-\ums{6}[M^BG_i^\dag F_k]^{ar}
\Big\}\,.
}
For the latter, if we simplify to the case of only one VLQ and one SM family, we obtain the result of 
Ref.\,\cite{kim.lewis}.

The operator $B\slashed{D}^3B$ operator can be simplified using the equations of motion \eqref{eom:B} and mostly gives 
$(M^B)^2/M^2$ corrections to lower dimensional operators or further loop suppressed contributions to tree-level 
contributions.
The only new operators generated are the SMEFT operators
\eq{
\lag_{\rm EFT}^{\text{1-$\ell$}}\supset
-\ums{4}H^\dag i\oLR{\slashed{D}}H \bar{q}_L\gamma^\mu (Y^BC_{BD^3B}{Y^B}^\dag) q_L
-\ums{4}H^\dag i\oLR{\slashed{D}}^I H \bar{q}_L\gamma^\mu\tau^I (Y^BC_{BD^3B}{Y^B}^\dag) q_L
\,.
}

This is a general consequence of the EFT power counting: dimension 6 operators are suppressed by $\Lcp^{-2}$ and any 
operator with dimension four generated by the use of the equations of motion in \eqref{eom:B} will be suppressed by the 
power $(M^B)^2/\Lcp^2$.
Operators with lower dimension will have a similar suppression and will be subleading compared to the directly 
generated 
ones. 
The only relevant use of the equations of motion \eqref{eom:B} is the reduction in one unit of operator dimension.
So the generation of a dimension five operator from a dimension six operator may induce an operator of similar order 
only suppressed by $M^B/\Lcp^2$.
And the only relevant ones are the dipole operators in \eqref{EFT:dim5:dipole}.

We omit the other dimension 6 operators. They include corrections to the tree-level ones in 
Table\,\ref{tab:op:mass:tree} and 
purely bosonic operators that are the same as when integrating out a singlet scalar added to the SM; see e.g. Table 3 
of 
Ref.\,\cite{cohen.zhang:20}.
We also have four-fermion operators of the SMEFT in the Warsaw basis replacing $d_R\to B_R$ and operators such as
\eq{
(H^\dag i\oLR{D}_\mu H)\bar{B}_L\gamma^\mu B_L
\,,
\quad
|H|^2\bar{B}_Li\slashed{D}B_L
\,,
\quad
(\bar{B}_L\sigma_{\mu\nu}B_R)^2
\,,
}
where the latter includes the variant with color indices swapped.

In summary, up to dimension 5, the only new operators generated at one-loop are the dipole operators in 
\eqref{EFT:dim5:dipole}.
See appendix \ref{ap:supertraces} for further discussion.

\section{Matching at $\Mvlq$}
\label{sec:Mvlq}

After integrating out the heavy scalar of the theory at $\Lcp$ and determining the effective theory with the presence 
of 
VLQs and SM fields (up to dimension 5), the next step is to integrate the VLQ out of the theory at $\Mvlq$ to find the 
SMEFT operators that are generated for this case. The integrating out at tree level of VLQs coupled to the SM 
with renormalizable couplings was performed in Ref.\,\cite{deBlas:2017xtg}.
Here, we are also considering additional dimension 5 operators that involve the VLQs and may also carry CP violation:
\eqali{
\label{lag:sm+vlq}
\lag^{\text{EFT}}_{\text{VLQ}}&=(\text{kinetic})
-(\bar{B}_LM^BB_R
+\bar{q}_LY^u\tilde{H} u_R+\bar{q}_LY^dHd_R+h.c.)
+\lag_{\rm VLQ}^{\rm int}
-V_H
\,,
}
where the VLQ interactions are
\eqali{
\label{lag:sm+vlq:int}
\lag^{\text{int}}_{\text{VLQ}}&=-\bar{q}_LHY^BB_R+\bar{B}_LC_{BdHH}d_R|H|^2
+\bar{B}_LC_{BBHH}B_R|H|^2
\\
&\quad 
+\bar{B}_LC^{\sigma B}_{Bd} \sigma^{\mu\nu}B_{\mu\nu} d_R
+\bar{B}_LC^{\sigma G}_{Bd} \sigma^{\mu\nu}T^A G^A_{\mu\nu} d_R
\\&\quad
+\bar{B}_LC^{\sigma B}_{BB} \sigma^{\mu\nu}B_{\mu\nu} B_R
+\bar{B}_LC^{\sigma G}_{BB} \sigma^{\mu\nu}T^A G^A_{\mu\nu} B_R
+h.c.,
}
and the Higgs potential is
\eq{
\label{higgs.pot}
V_H=\lambda\big(|H|^2-v^2/2\big)^2+\cdots\,.
}
The ellipsis may include operators such as $|H|^6$ that are generated at $\Lcp$ and are listed in 
Table~\ref{tab:op:mass:tree}.
We denote the Lagrangian \eqref{lag:sm+vlq} as the VLQ-EFT and we adopt the convention where $\arg\det M^B=0$.
The two parameters with positive mass dimension are $M^B$ and $v^2$, where the latter is traded 
with\,\cite{jenkins.manohar:1}
\eq{
\label{def:mH}
m^2_H\equiv 2\lambda v^2\,.
}
With tree-level matching, the potential \eqref{higgs.pot} is indeed the Higgs potential of the SM with 
$v\approx 246\,\unit{GeV}$ and \eqref{def:mH} is the Higgs boson mass.
Note that here the couplings and Wilson coefficients are general and they need not coincide with
the coefficients coming from the matching at $\Lcp$ with the same name.
For example, for the theory matched at one-loop, one should consider the effective couplings in 
\eqref{lag:1-L:redefined}.
As another example, from the matching at $\Lcp$ without running, the Wilson coefficients $C^{\sigma B}$ and $C^{\sigma 
G}$ would be related by \eqref{F.sigma} and \eqref{dipole.eff} as
\eqali{
\label{dipole:eft.eff}
C^{\sigma B}_{Bd}&=-\ums{3}g'C^{\sigma\rm eff}_{Bd}\,,\quad
C^{\sigma G}_{Bd}&=g_s C^{\sigma\rm eff}_{Bd}\,,
\cr
C^{\sigma B}_{BB}&=-\ums{3}g'C^{\sigma\rm eff}_{BB}\,,\quad
C^{\sigma G}_{BB}&=g_s C^{\sigma\rm eff}_{BB}\,.
}

We perform below the matching at tree-level and one-loop.
For the latter we focus only on the threshold corrections that affect $\btheta$.

\subsection{Tree level matching at $\Mvlq$}
\label{sec:tree:vlq}

With the kinetic terms for the VLQs 
\eq{
\bar{B}_Li\slashed{D}B_L+\bar{B}_Ri\slashed{D}B_R-(\bar{B}_LM^BB_R+h.c.)\,,
}
and $\lag^{\text{int}}_{\text{VLQ}}$, we can find the classical fields for the VLQs from the EOMs:
\eqali{
B_R&=\big(M^B\big)^{-1}\left(i\slashed{D}B_L+\frac{\delta \lag^{\text{int}}_{\text{VLQ}}}{\delta \bar{B}_L }\right),\\
B_L&=\big(M^{B\dag}\big)^{-1}\left(i\slashed{D}B_R+\frac{\delta \lag^{\text{int}}_{\text{VLQ}}}{\delta \bar{B}_R 
}\right),
}
which must be employed iteratively.
For example, $B_L\sim (M^B)^{-1}\text{(dim.\,5/2)}$ and $B_R\sim (M^B)^{-2}\text{(dim.\,7/2)}$ at leading order.
Following Ref.\,\cite{Criado:2019mvu}, we can identify the operators generated at tree level in SMEFT, along with their
corresponding Wilson coefficients, as summarized in Table~\ref{tab:op:vlq:tree}.
These Wilson coefficients match the results of Ref.\,\cite{Criado:2019mvu} if we only consider the third family.
\begin{table}[ht]
\[
\begin{array}{|c|c|c|c|c|}
\hline
 & \text{Operator} &  \text{Wilson coefficient}&\text{Power-counting for NB
 }
 \\
\hline
\rule{0ex}{3ex}
\mathcal{O}^{(1)}_{Hq} & (H^\dag i\oLR{D}_\mu H)(\bar{q}_L\gamma^\mu q_L) &  
  -\ums{4}G_B
  &  \left(M^B\right)^{-2}
\\[.5ex]
\mathcal{O}^{(3)}_{Hq}
&(H^\dag i\oLR{D}^I_\mu H)(\bar{q}_L\gamma^\mu \tau^I q_L)   
  & -\ums{4}G_B
  &\left(M^B\right)^{-2}
\\[.5ex]
\mathcal{O}_{dH} &\bar{q}_L Hd_R|H|^2  
  &   
  \ums{2}G_BY^d
  -Y^B(M^B)^{-1}C_{BdHH}
  &\left(M^B\right)^{-2}+\left(M^B\right)^{-1}\Lcp^{-1}
\\[.5ex]
\mathcal{O}_{dB} &\left(\bar{q}_L\sigma^{\mu\nu} d_R\right)HB_{\mu\nu}  &  - Y^B( M^B)^{-1}\left(C^{\sigma B 
}_{Bd}\right)&\left(16\pi^2\Lcp^2\right)^{-1}
\\[.5ex]
\mathcal{O}_{dG} &\left(\bar{q}_L\sigma^{\mu\nu}T^A d_R\right)HG^A_{\mu\nu}  & -  Y^B( M^B)^{-1}\left(C^{\sigma 
G}_{Bd}\right)&\left(16\pi^2\Lcp^2\right)^{-1}
\\[.7ex]
\hline
\end{array}
\]
\caption{\label{tab:op:vlq:tree}%
Operators and Wilson coefficients in SMEFT in Warsaw basis after integrating out the VLQs at tree level.
For non-hermitian operators, the presence of their hermitean conjugates is understood.
The operators follow the convention of Ref.\,\cite{Jenkins:2017jig} with the two quark flavor indices implicit
in matrix notation and $G_B$ is defined in \eqref{def:GB}.
}
\end{table}

It is known that with VLQs coupling to the SM only through renormalizable interactions, 
i.e., the first term of \eqref{lag:sm+vlq:int},
only three dimension 6 operators are generated at tree level\,\cite{deBlas:2017xtg}.
They correspond to the first three operators in Table~\ref{tab:op:vlq:tree}.
For notational simplicity we have defined
\eq{
\label{def:GB}
G_B\equiv Y^B({M^B}^\dag M^B)^{-1}{Y^B}^\dag\,.
}
When we add the dimension 5 operators in the VLQ-EFT of \eqref{lag:sm+vlq:int},
we gain two more operators in the SMEFT corresponding to dipole operators.\footnote{%
There are additional contributions to dipole operators from one-loop matching at $\Mvlq$
that do not contribute to CP violation. They are briefly discussed in Sec.\,\ref{sec:results}.
}
The Higgs potential \eqref{higgs.pot} remains the same at tree-level matching.

\subsection{One-loop matching at $\Mvlq$}

The integration of heavy fermion loops at one-loop level has already been discussed in 
Refs.\,\cite{Huo:2015exa,fermion:UOLEA}, with universal formulas but without considering light-heavy loops.
The latter can be included with clever block-diagonalization\,\cite{Kramer:2019fwz} after some adaptation.
All these formulas are applicable to renormalizable theories which is not the case of our theory, the VLQ-EFT in 
\eqref{lag:sm+vlq}.
As such, we do not perform the full one-loop matching and focus on the consequences to $\btheta$
following mainly Ref.\,\cite{cohen.zhang:20}.
In the same reference one can track the CP even operators of dimension four, depending purely on gauge 
bosons, that are universally generated from the log-type term in \eqref{log.power}.

Our UV theory for one-loop matching is the VLQ-EFT in \eqref{lag:sm+vlq}.
Prior to computing the one-loop functional matching, it is necessary to determine the $U$ matrices arising from the 
interactions between the heavy field and the light fields in the theory. In this section, the heavy field corresponds 
to the VLQ, while the light fields are those of the SM.
The relevant $U$ matrices are
\eq{
\begin{aligned}
\label{Mvlq:U}
U_{qB}^{[1]}&=\begin{pmatrix}
Y^B RH&0\\
0&{Y^{B}}^* LH^*
\end{pmatrix},
\qquad
U_{Bd}^{[2]}=
-
\begin{pmatrix}
C_{BdHH}|H|^2 R&0\\
0&C^{*}_{BdHH}|H|^2L
\end{pmatrix},
\cr
U_{HB}^{[\nicefrac{3}{2}]}&=
\mtrx{0& \bar{q}^c L{Y^{B}}^*\cr \bar{q}R{Y^{B}}&0}
\cr
&\quad -
\begin{pmatrix}
H\big(\bar{B}RC_{BBHH}+\bar{B}LC^\dag_{BBHH}+\bar{d}LC^\dag_{BdHH}\big) 
&
H\big(\bar{B}^cLC^*_{BBHH}+\bar{B}^cRC^\tp_{BBHH}+\bar{d}^cRC^\tp_{BdHH}\big)
\\
H^*\big(\bar{B}RC_{BdHH}+\bar{B}LC^\dag_{BBHH}+\bar{d}LC^\dag_{BdHH}\big)
&
H^*\big(\bar{B}^cLC^*_{BBHH}+\bar{B}^cRC^\tp_{BBHH}+\bar{d}^cRC^\tp_{BdHH}\big)
\end{pmatrix},
\cr
U_{BB}^{[2]}&=
-|H|^2\mtrx{
C_{BBHH}R+C^\dag_{BBHH}L & 0 \cr 
  0 & C_{BBHH}^*L+C^\tp_{BBHH}R
  }\,.
\end{aligned}
}
For the same reason as in \eqref{U:2}, the $SU(2)_L$ indices of the doublet $\bar{q}$ are understood as a 
column vector.

The one-loop threshold contributions that will be relevant for $\bar{\theta}$ are
\eqali{
\label{lag:1l:dim4:VLQ}
\lag_{\rm EFT}^{\text{1-$\ell$}} & \supset\delta Z_H(D_\mu H)^\dagger(D^\mu H)
+\bar{q}_L\delta Z_{q_L}i\slashed{D}q_L
\\
&
+\Big(\bar{q}_{L}\delta Y^d_GHd_{R}+\bar{q}_{L}\delta Y^u_G\tilde{H}u_{R}+h.c.\Big)
}
The contribution to the Higgs kinetic term and to that of the quark doublets are equal to
\eqali{
\label{deltaZ:qL H:eff}
16\pi^2\delta Z_H&=\Big\langle{Y^B}^\dag Y^B \left(\ums{2}\id_n+L_B\right)\Big\rangle\,,\\
16\pi^2[\delta Z_{q_L}]^{pr}&=
\ums{2}\big[Y^B\left(\ums[3]{2}\id_n+L_B\right){Y^B}^\dag\big]^{pr}\,,
}
where $\aver{~~}$ denotes a trace if there are more than one VLQ and
$L_B$ is
\eq{
[\hat{L}_B]^{ab}\equiv \delta_{ab}\log\frac{\mu^2}{M^2_{B_a}}\,,
}
in the basis where $M^B$ is diagonal and $\mu=\Mvlq\sim M_{B_a}$.
When $M^B$ is nondiagonal, we define two generalizations for the diagonal $L_B$ as
\eq{
L_B=-\log[{M^B}^\dag M^B/\mu^2]\,,\quad
\tL_B=-\log[M^B{M^B}^\dag /\mu^2]\,.
}
In eq.\,\eqref{deltaZ:qL H:eff}, $L_B$ is understood as in the first expression.

\begin{figure}
\includegraphics[scale=.75]{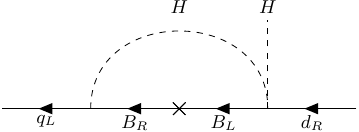}
\caption{\label{diag:deltaYdCbdHH}%
One-loop contribution to $Y^d_{\rm eff}$ in \eqref{deltaYd:G:CBdHH} relevant to $\btheta$ when integrating out the 
VLQs. 
}
\end{figure}

The corrections to the Yukawa couplings are
\subeqali{
\label{deltaYd:G:CBdHH}
16\pi^2[\delta Y^d_G]^{pr}&=-[Y^B\left(\id_n+L_B\right){M^B}^{\dagger}C_{BdHH}]^{pr}\,,
\\
\label{deltaYu:G}
16\pi^2[\delta Y^u_G]^{pr}&=-[Y^B\left(\id_n+L_B\right){Y^B}^\dagger Y^u]^{pr}\,.
}
For $Y^d$, note the contribution from the five dimensional operator $\bar{B}d|H|^2$ in \eqref{lag:sm+vlq:int}
which is depicted in Fig.\,\ref{diag:deltaYdCbdHH}.
The correction to $\delta Y^d_G$ comes from $U_{BH}U_{HB}$ while the one to $\delta Y^B_G$ comes from 
$U_{BH}U_{Hq}U_{qB}$.

After field redefinitions, similar to the ones in Sec.\,\ref{Lcp:field.redef},
the effective Yukawa couplings in the SMEFT are
\eqali{
[Y^u_{\rm eff}]^{pr}&=[Y^u(1-\ums{2}\delta Z_{H})-\ums{2}\delta Z_{q_L}Y^u-\delta Y^u_G]^{pr}\,,
\\
[Y^d_{\rm eff}]^{pr}&=[Y^d(1-\ums{2}\delta Z_{H})-\ums{2}\delta Z_{q_L}Y^d-\delta Y^d_G]^{pr}\,.
}
As the corrections for $Y^u$ are all hermitean matrices multiplying $Y^u$, they do not contribute to
$\btheta$.
Similarly, for $Y^d$, only the piece involving $C_{BdHH}$ will contribute.
Therefore, the threshold contribution at $\Mvlq$ is
\eq{
\label{theta:CbdHH}
16\pi^2\delta\btheta(\Mvlq)=
-
\im\tr\big[
{Y^d}^{-1}Y^B\left(\id_n+L_B\right){M^B}^{\dagger}C_{BdHH}\big]
\,;
}
see Sec.\,\ref{sec:Lcp:theta-bar} for our definition of $\btheta$.
So far, this is a completely general result for the matching between VLQ-EFT in \eqref{lag:sm+vlq} 
and the SMEFT and is not restricted to Nelson-Barr theories.
As such, it should be considered as an one-loop correction in the matching at $\Mvlq$
\emph{additional} to any contribution that comes from the tree-level $Y^d$ or $M^B$.

For Nelson-Barr theories, $C_{BdHH}$ is determined from other couplings in the UV complete theory with
the CP breaking scalars as in Table \ref{tab:op:mass:tree}.
Neglecting the running between $\Lcp$ and $\Mvlq$, we can rewrite \eqref{theta:CbdHH} in terms of the
couplings of the UV complete theory \eqref{lag:UV:mass} as
\eqali{
\label{theta:vlq:uv}
16\pi^2\delta\btheta(\Mvlq)&=
-(M^{-2}\gamma')_i\im\tr[F_i{Y^d}^{-1}Y^B(\id_n+L_B){M^B}^\dag]\,,
\cr
&=-(M^{-2}\gamma')_i\im\tr[\cF_i{\cM^{Bd}}^\dag(\id_n+\tL_B) ]\,,
}
where the first expression is in the VLQ mass basis while the last is in the CP basis. 
For the latter, the mass matrix $M^B{M^B}^\dag$ is generically nondiagonal.
These expressions are very similar to \eqref{thetabar:int.si} and \eqref{thetabar:int.si:cp}.

\subsection{Running between $\Lcp$ and $\Mvlq$}

We have calculated in Eqs.\,\eqref{thetabar:int.si} and \eqref{theta:vlq:uv} the finite threshold corrections
to $\btheta$ in Nelson-Barr theories when integrating out, respectively, the CP breaking scalars at $\Lcp$ and 
integrating out the VLQs at $\Mvlq$.
The matching at one-loop were performed between the UV complete theory \eqref{lag:UV:mass}
and the VLQ-EFT \eqref{lag:sm+vlq}, and subsequently between the VLQ-EFT and the SM (SMEFT).
These corrections are very similar in form.

However, when the separation between $\Lcp$ and $\Mvlq$ is large, RGE running effects may affect 
substantially the results. This is what we discuss here.

We focus on the running of $\btheta$ in the VLQ-EFT \eqref{lag:sm+vlq}.
Within the SM, the running of $\btheta$ is negligible as it is expected to arise only at seven 
loops\,\cite{ellis.gaillard,vainshtein}.
For generic renormalizable theories the running of $\theta$ is expected to arise only at two loops
and the Yukawa contributions to $\btheta$ may run at one loop only for special models with at least 
two scalars in different representations of QCD\,\cite{vecchi:theta}. This is not the case of the renormalizable
part of any model we are considering.
Therefore, for our purposes, it is sufficient to track only the contributions to the running of $\btheta$ with
the participation of at least one of the dimension five operators in the VLQ-EFT.

\begin{figure}[h]
\includegraphics[scale=.8]{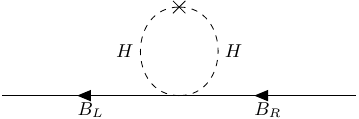}
\caption{\label{diag:deltaMB-CBBHH}%
One-loop contribution to $\delta M^B$ in \eqref{deltaMB:G:CBBHH} contributing to the running of $\btheta$
in the VLQ-EFT \eqref{lag:sm+vlq:int}.
}
\end{figure}

For the extraction of the RGE, we employ the method of Ref.\,\cite{Henning:2016lyp} which calculates
the effective action through CDE and imposes the constancy of the effective coupling of each operator after
rescaling the kinetic parts to canonical form.
Since the contribution from the wave-function corrections are hermitean corrections multiplying the original Yukawas, 
they do not contribute to the imaginary part and can be neglected.
In this case, differently from matching to an EFT, there is no expansion on the ``light'' fields.
The relevant corrections are 
\subeqali[running:correction]{
16\pi^2[\delta Y^d_G]^{pr}&\supset -[Y^B\left(\id_n+L_B\right){M^B}^{\dagger}C_{BdHH}]^{pr}\,,
\\
\label{deltaMB:G:CBBHH}
16\pi^2[\delta M^B_G]^{pr}&\supset m_H^2[C_{BBHH}]^{pr}\Big(1+\log\frac{\mu^2}{m^2_{\rm IR}}\Big)
\,,\\
16\pi^2[\delta C^{Bd}]^{pr}&
\supset m_H^2[C_{BdHH}]^{pr}\Big(1+\log\frac{\mu^2}{m^2_{\rm IR}}\Big),
}
where $m_H$ was defined in \eqref{def:mH} and $m_{\rm IR}$ is an IR regulator.
The notation is similar to \eqref{lag:1l:dim4:VLQ}.
The VLQ-EFT \eqref{lag:sm+vlq} possesses two dimensionful parameters that can appear on the numerator: $M^B$ and
$m_H$ (or $v$). The latter is the only dimensionful parameter in the SMEFT and it appears in the running of
some dimensionless parameters of the SM, e.g., in the contribution of $C_{uH}$ to the running of
$Y^u$\,\cite{jenkins.manohar:1}.
The first correction in \eqref{running:correction} is identical to \eqref{deltaYd:G:CBdHH} which we just repeat.
The second correction involves a Higgs loop and is depicted in Fig.\,\ref{diag:deltaMB-CBBHH}.
The third correction is similar after exchanging $B_R$ by $d_R$.
As the loop involves a light field, the calculation involves the additional $U$ matrix
\eqali{
U'_{H H}=&-\Big\{\bar{B}_L\big(C_{BBHH}B_R+C_{BdHH}d_R\big)+h.c.\Big\}\left(\begin{array}{ll}
\mathbb{1} & 0 \\
0 & \mathbb{1}
\end{array}\right)
\,.
}

Considering the contributions \eqref{theta.d.B}
to $\btheta$ in \eqref{theta:Lcp} which is valid
within the VLQ-EFT, the RGE for $\btheta$ is\,\footnote{%
We neglect the one-loop running of $M^B,C^{Bd},Y^d,Y^B$ induced by the dipole operator 
$C^{\sigma G}_{Bd}$ which in the NB setting gives a contribution further 
suppressed by $(g_s/4\pi)^2\Mvlq^2/\Lcp^2$ compared to the first contribution in \eqref{theta:rge}, ignoring 
${Y^d}^{-1}Y^B$.
The contribution from $C^{\sigma B}_{Bd}$ and the analogs replacing $d_R$ by $B_R$ are similarly suppressed. 
A similarly suppressed matching contribution would be added to \eqref{deltaYd:G:CBdHH} as well.
}
\eqali{
\label{theta:rge}
16\pi^2\mu \frac{d}{d \mu}\btheta&=-2\im\tr\left[{Y^{d}}^{-1}Y^B{M^B}^{\dagger}C_{BdHH}\right]
\cr&\quad
+2m_H^2\im\tr\left[{M^B}^{-1}\big(C_{BBHH}-C_{BdHH}{Y^{d}}^{-1}Y^B\big)\right]\,.
}
This result is valid independently if the original UV complete theory is a Nelson-Barr theory.
For a Nelson-Barr theory, the first term in \eqref{theta:rge} scale as $(M^B/\Lcp)^2$
while the second term proportional to $m^2_H$ \emph{vanishes} due to \eqref{BBHH->BdHH}.

The VLQ-EFT \eqref{lag:sm+vlq:int} is thus an example of a simple (non-renormalizable) BSM theory where $\btheta$ runs 
already at one-loop through the presence of dimension five operators.
Given the strong constraint on $\btheta$, this may put strong restrictions on the Wilson coefficients of these 
operators, specially on $C_{BdHH}$ in this case.

\subsection{Broken phase of SMEFT}

It is well known that the SMEFT with dimension six operators loses the characteristic property (at tree-level) of 
the SM in which both the fermion mass matrices and the fermion-Higgs couplings are proportional to the respective 
Yukawa matrices\,\cite{buchmuller.wyler,jenkins.manohar:3}.
For a down-type singlet VLQ, the induced operator $\cO_{dH}=\bar{q}_LHd_R|H|^2$ modifies the relation between the 
down quark mass matrix $M_d$ and the Yukawa coupling $Y^d$ as
\eq{
\label{smeft:Md}
[M^d]^{rs}=\frac{v_T}{\sqrt{2}}\bigg([Y^d]^{rs}-\ums{2}v^2[C_{dH}]^{rs}\bigg)
\,.
}
In the context of the strong CP problem, the second term leads to an additional contribution to $\btheta$ at \emph{tree 
level}.

Before analyzing the SMEFT case, let us discuss the case of the VLQ-EFT \eqref{lag:sm+vlq:int}.
When the VLQs lie hypothetically at the electroweak scale, there is no need to integrate them out and the full mass 
matrix of down-type quarks after the Higgs gains the vev $v$ is of size $(3+n)$:
\eq{
\cM^{d+B}=\mtrx{\ums{\sqrt{2}}Y^d v & \ums{\sqrt{2}}Y^B v
\cr
-\ums{2}C_{BdHH}v^2 & M^B-\ums{2}C_{BBHH}v^2
}\,.
}
In the NB theory, the Wilson coefficients above are related by \eqref{BBHH->BdHH} and one can check that
\eq{
\arg\det\cM^{d+B}=0\,,
}
at linear order in the Wilson coefficients considering $\arg\det Y^d=\arg\det M^B=0$.
This leads to the correct result that $\btheta=0$ \emph{at tree level} in the NB construction.

Now, if we integrate the VLQs out of theory at \emph{tree-level} as in Sec.\,\ref{sec:tree:vlq}, the coefficient 
$C_{BdHH}$ contributes 
to $\btheta$ through \eqref{smeft:Md} but the compensating contribution of $C_{BBHH}$ disappears 
from the SMEFT up to dimension six.
We end up with a nonzero contribution to $\btheta$ at \emph{tree level}, conflicting with the original Nelson-Barr 
UV theory.

The resolution to this conundrum is that a \emph{one-loop} contribution to the operator $|H|^2G\tilde{G}$ operator 
yields a contribution to $\theta$ equivalent to \emph{tree-level} (without $16\pi^2$ suppression) when $H$ gets a vev.
The actual calculation in appendix \ref{ap:UBB} yields\,\footnote{%
There are additional contributions to $|H|^2G\tilde{G}$ involving the insertion of the dipole operators 
\eqref{EFT:dim5:dipole} when integrating out the VLQ. Their contribution to $\btheta$ goes as  
$(16\pi^2)^{-1}(v^2/\Lcp^2)(Y^2/g_s)$ where $Y^2=(Y^B)^2,(Y^BY^d)$ and they are negligible compared to the leading 
$\Mvlq/\Lcp$ contributions discussed here.
}
\eqali{
\label{HHGG-tilde}
\lag_{\rm SMEFT} &\supset 
\im\Tr[{M^B}^{-1}C_{BBHH}]|H|^2 
\frac{g_s^2}{64\pi^2}\epsilon^{\mu\nu\alpha\beta}G^a_{\mu\nu}G^a_{\alpha\beta}
\,.
}
This contribution comes from the supertrace containing one insertion of $U_{BB}$ in \eqref{Mvlq:U}.
When $H$ gets the vev, this contributes to the QCD $\theta$ term with
\eq{
\label{thetaB'}
\theta_B'=\frac{v^2}{2}\im\Tr\Big[{M^B}^{-1}C_{BBHH}\Big]\,.
}
Comparing to \eqref{smeft:Md}, we can see that the contribution from the operator $\cO_{dH}=\bar{q}_LHd_R|H|^2$ in the 
SMEFT yields a contribution additional to $\btheta=-\arg\det Y^d$ as
\eq{
\label{theta:Mvlq:Md}
\theta_d'=\frac{v^2}{2}\im\Tr[{Y^d}^{-1}C_{dH}]
\,.
}
Within the SMEFT, this contribution is generic once in the broken phase 
and an analogous contribution may also arise from the up quark mass matrix.

In NB theories, the coefficient $C_{dH}$ is given in Table\,\ref{tab:op:vlq:tree} and the nontrivial contribution to 
$\theta_d'$ comes solely from $C_{BdHH}$.
The relation \eqref{BBHH->BdHH} finally implies that for the Nelson-Barr theory 
matched to the SMEFT, we correctly get the expected result
\eq{
\btheta(v)=\theta_B'+\theta_d'=0\,,
}
neglecting the running between $\Mvlq$ and $v$.
We should emphasize once more that $\theta_d'$ comes from \emph{tree-level} matching while $\theta_B'$ comes from 
\emph{one-loop} matching.
In theories with VLQs that do not descend from NB theories, this cancellation would not occur and each contribution 
would be subjected to strong contraints from $\btheta$.

\subsection{Running in SMEFT}

The new operators in the SMEFT introduced at the matching at $\Lcp$ are listed in Table \ref{tab:op:mass:tree} 
(tree-level) and equation \eqref{EFT:dim5:dipole} (one-loop up to dimension 5).
The new operators introduced at the matching at $\Mvlq$ are listed in Table\,\ref{tab:op:vlq:tree} (tree-level at 
$\Mvlq$).
Here we discuss the running of $\btheta$ within the SMEFT originating from the RGEs of the SM couplings (dimension four 
operators) induced by dimension six operators\,\cite{jenkins.manohar:1}.

In this context, two sources enter in the running of $\btheta$: the running of Yukawas $Y^d,Y^u$ and the running 
of the topological angle $\theta$.
The relevant operators that enter in the RGEs of $Y^u,Y^d$ are $\cO_{H\Box},\cO_{Hq}^{(1)},\cO_{Hq}^{(3)},\cO_{dH}$ 
(for up-type VLQ the latter is $\cO_{uH}$).
But almost all of them are proportional to the Yukawa themselves leaving only a hermitean matrix with real trace.
They do not contribute to $\btheta$.
We are only left with the contribution from the running of $Y^d$ induced by $\cO_{dH}$ which leads to
\eq{
16\pi^2\mu\frac{d}{d \mu}\btheta\Big|_{Y^d}=-3m_H^2\im\Tr[{Y^d}^{-1}C_{dH}]\,.
}
This equation can be easily adapted for up-type VLQs.
Writing $C_{dH}$ in Table\,\ref{tab:op:vlq:tree} in terms of the couplings of the VLQ-EFT 
interactions \eqref{lag:sm+vlq:int}, we 
obtain
\eq{
\label{rge:theta:smeft}
16\pi^2\mu\frac{d}{d \mu}\btheta\Big|_{Y^d}=3m_H^2\im\Tr\big[{Y^d}^{-1}Y^B{M^B}^{-1}C_{BdHH}\big]\,.
}

The running of the topological angle $\theta$ itself is induced by the operator 
$|H|^2G\tilde{G}$\,\cite{jenkins.manohar:1}:
\eq{
\label{rge:theta}
\mu\frac{d}{d\mu}\theta=+\frac{4m_H^2}{g_s^2}C_{H\tilde{G}}\,,
}
where we adapted the sign of $\theta$ to our convention in Sec.\,\ref{sec:Lcp:theta-bar}
and the Wilson coefficient $C_{H\tilde G}$ can be extracted from \eqref{HHGG-tilde}
as\,\footnote{
Ref.\,\cite{jenkins.manohar:1} uses the Warsaw basis convention\,\cite{warsaw} where 
$\cO_{H\tilde G}=-|H|^2\ums{2}\epsilon^{\mu\nu\alpha\beta}G^a_{\mu\nu}G^a_{\alpha\beta}$
once adapted to our convention.
}
\eq{
C_{H\tilde G}=-\im\Tr[{M^B}^{-1}C_{BBHH}]\frac{g_s^2}{32\pi^2}
\,.
}
Adding the contribution \eqref{rge:theta:smeft}, the full RGE for $\btheta$ within the SMEFT is
\eq{
16\pi^2\mu\frac{d}{d\mu}\btheta=
3m_H^2\im\Tr\big[{Y^d}^{-1}Y^B{M^B}^{-1}C_{BdHH}\big]
-2m_H^2\im\Tr[{M^B}^{-1}C_{BBHH}]\,.
}
We can see that in the NB theory the relation \eqref{BBHH->BdHH} implies only a partial cancellation
and the one-loop running \emph{does} occur.
In leading log approximation, the \emph{additional} contribution from $\Mvlq$ down to $v$ is
\eq{
\delta\btheta(v)=\frac{m_H^2}{16\pi^2}
\Big\{
3\im\Tr\big[{Y^d}^{-1}Y^B{M^B}^{-1}C_{BdHH}\big]
-2\im\Tr[{M^B}^{-1}C_{BBHH}]
\Big\}
\log\frac{v}{\Mvlq}\,.
}

Therefore, $\btheta$ already runs at one-loop in SMEFT in the presence of some dimension six operators.
This running may be relevant if the new physics scale, $\Mvlq$ in our case, is very far from the electroweak scale.

Below the electroweak scale, we can track the running in the low energy EFT (LEFT) where 
$t, h, W, Z$ are integrated out.
At one-loop the only contribution comes from the dimension five dipole operators that comes from the dimension six 
dipole operators in the SMEFT (see eq.(C.7) in Ref.\,\cite{Jenkins:2017dyc}).
In the NB case, the dipole operators with imaginary parts have coefficients scaling as $1/(16\pi^2\Lcp^2)$
which  induces the running $\mu\frac{d}{d\mu}\btheta\sim vm_b/(16\pi^2\Lcp^2)\sim 10^{-16}$ in LEFT.
This is quite negligible.

\section{Results and applications}
\label{sec:results}

In Nelson-Barr theories, VLQs should transmit the CP breaking from the (spontaneous) CP breaking sector to the CP 
conserving version of the SM.
Reproducing the CP violation of the SM in the CKM mixing on the one hand and avoiding massless SM quarks on the other 
hand requires\,\footnote{%
It is possible to consider $|\cM^{Bd}|\gg |\cM^B|$ to some degree, a regime that was denoted as the \emph{seesaw limit} 
in Ref.\,\cite{nb-vlq:seesaw} which helps explaining some hierarchies of the SM Yukawas.
For the actual possible ranges for $|{\cM^B}^{-1}\cM^{Bd}|$, see Refs.\,\cite{nb-vlq,nb-vlq:more}.
}
\eq{
\cM^{Bd}\sim \cM^B\sim M^B\sim \Mvlq\,.
}
As $\cM^{Bd}=\cF_iu_i$ is connected to the CP breaking vevs $u_i\sim \Lcp$, perturbativity also demands that 
$\cM^{Bd}\lesssim \Lcp$.
To be able to observe some new states, we will assume that $\cM^{Bd}\ll \Lcp$ and we will see below that this hierarchy
automatically suppresses the one-loop corrections to $\btheta$ to acceptable values.
Although the weak scale needs to be stabilized by other means\,\cite{BBP}, a low scale for $\Mvlq$ is technically 
natural in the Nelson-Barr scheme.

For the Nelson-Barr theory our main results for the contribution for $\btheta$ are given in
equations \eqref{thetabar:int.si} (one-loop threshold contribution at $\Lcp$), \eqref{theta:CbdHH}
(one-loop threshold correction at $\Mvlq$) and \eqref{theta:rge} (one-loop RGE in VLQ-EFT).
With one-loop matching at $\Lcp$, running down $\btheta$ in the leading log approximation
and one-loop matching at $\Mvlq$, we get
\eqali{
\label{theta:Mvlq:sum}
16\pi^2\btheta(\Mvlq)&=\text{Eq.\,\eqref{thetabar:int.si}}
+\text{Eq.\,\eqref{theta:CbdHH}}
\cr
&\quad
-
2\im\tr\left[{Y^{d}}^{-1}Y^B{M^B}^{\dagger}C_{BdHH}\right]
\log\frac{\Mvlq}{\Lcp}
\,.
}
If we use tree-level matching, we can disregard the first line
and the one-loop running \eqref{theta:rge} is the most relevant contribution that needs to be traced.
We can also rewrite \eqref{theta:Mvlq:sum} in terms of the couplings in the UV complete theory in the CP basis
\eqref{lag:UV:cp} as in eq.\,\eqref{theta:vlq:uv}.
Keeping only the dominant log term, using \eqref{MBd:ui} and $\gamma'_i=\gamma_{ij}u_j$, we obtain
\eq{
\label{theta:NB:log}
\btheta(\Mvlq)=-\frac{1}{8\pi^2}
u_l\gamma_{lj}M^{-2}_{ji}u_k\im\tr\left[\cF_i\cF_k^\dag\right]
\log\frac{\Mvlq}{\Lcp}
\,.
}
Imposing $|\btheta|< 10^{-10}$, we obtain
\eq{
\Big|u_l\gamma_{lj}M^{-2}_{ji}u_k\im\tr\left[\cF_i\cF_k^\dag\right]\Big|<8\times 10^{-10}
\frac{10}{\log\frac{\Lcp}{\Mvlq}}\,,
}
where the lefthandside is independent of the scale $\Lcp$.
The essential UV parameters are the VLQ Yukawas $\cF_i$ and Higgs portal couplings $\gamma_{ij}$, both to the CP 
breaking scalars $s_i$.
The expression above is the same considered in Ref.\,\cite{dine} without the log enhancement
which is present in Refs.\,\cite{reece,murai.nakayama:1}.
The log enhancement is relevant for a large separation between $\Lcp$ and $\Mvlq$.
For example, $\log(10^8\,\unit{GeV}/1\,\unit{TeV})\approx 11.5$.
Note that the log in Refs.\,\cite{reece,murai.nakayama:1} depends on the ratio $\Lcp/v$ while 
in \eqref{theta:NB:log} the log depends on the ratio between $\Lcp$ and $\Mvlq$ and should be continued within the 
SMEFT, cf.\,\eqref{rge:theta:smeft}, if $\Mvlq$ is very far from the electroweak scale.

In fact, in Nelson-Barr theories, the presence of new physics at say $\Mpl$ would induce the dimension five
operators $s_is_j\bar{B}_LB_R$ and $s_i\bar{q}_LHB_R$ which would spoil
the Nelson-Barr structure and
induce too large contributions to $\btheta$ unless $\Lcp$ is relatively below $\Mpl$\,\cite{dine,reece} as
\eq{
\Lcp\lesssim 10^8\,\unit{GeV}\,.
}
This is the so-called \emph{quality} problem in Nelson-Barr theories.
On the other hand, lowering too much the scale $\Lcp$ associated to the spontaneous breaking of CP leads to
cosmological problems.
For example, if CP is a gauge symmetry ---expected from the nonconservation of global symmetries by gravity---
it is argued\,\cite{reece:stable} that the domain wall from spontaneous CP breaking is absolutely stable and 
needs to be inflated away.
This requires inflation to take place after spontaneous CP breaking and $T_{\rm reh}<\Lcp$.\footnote{%
This might be circumvented by inverse symmetry breaking\,\cite{murai.nakayama:1}.
}

If we consider that $\Lcp$ is not so much above the electroweak scale but still $\Lcp\gg v$, we can consider the UV 
theory given by the potentials \eqref{VS:general}  and \eqref{VHS:general}.
If we consider $s_1$ with mass $M_1$ to be much lighter than the rest, $\gamma_1'$ will contribute to the mixing with 
the SM Higgs with mixing angle $\sin\alpha\sim v\gamma'_1/M_1^2$. This mixing is currently constrained as 
$|\sin\alpha|\lesssim 0.2$ by Higgs coupling deviations and precision electroweak observables\,\cite{Dawson:2021}, and 
for $\gamma_1'\sim \gamma_1\Lcp$, $\gamma_1$ is already constrained more by the perturbativity limit $4\pi$ for 
$M_1\sim \Lcp\gtrsim 15\,\unit{TeV}$ than this mixing value.

As for VLQs, the most stringent collider limit for singlet VLQs $B$ or $T$ currently comes from CMS searches through 
pair production:
\eq{
m_T>1.48\,\unit{TeV}\,\text{\cite{cms}}\,,\quad
m_B>1.49\,\unit{TeV}\,\text{\cite{cms.review}}\,.
}
We will consider the lower limit $1.48\,\unit{TeV}$ for both cases.
These searches assume that the VLQ couples exclusively to third family SM quarks which is approximately the case for 
NB-VLQs, cf.\,\eqref{typical.YB}.
For a summary of other collider constraints see Refs.\,\cite{benbrik,vlq-review,cms.review}, which includes 
single production searches relevant for large VLQ masses.
Generically, precision electroweak observables constrain the mixing of VLQs to SM quarks which scales as 
$Y^B/M^B$\,\cite{handbook}.
Generic Yukawa couplings $Y^B$ are also strongly constrained by flavor physics\,\cite{buras.celis.17,ligeti.wise} but 
for NB-VLQs following the hierarchical structure \eqref{typical.YB}, the constraints are 
similar\,\cite{nb-vlq:fit,nb-vlq:more,vecchi.1} for TeV scale VLQs.
In contrast, constraints for the Wilson coefficient for the dimension six SMEFT operators such as 
$\cO_{H\Box},\cO_{Hq}^{(1)},\cO_{Hq}^{(3)}$ that are generated are weak\,\cite{ellis:smeft} and are not relevant for a 
high effective scale $\Lcp$.
Analogously, the current constraints for the dimension five operators of the VLQ-EFT are equally not 
relevant\,\cite{Criado:2019mvu}.

Focusing again on $\btheta$, we can obtain a constraint in a different form by rewriting \eqref{theta:NB:log} in terms 
of the quantities in the VLQ
mass basis \eqref{lag:UV:mass}
as
\eq{
\label{theta:NB:log:MB}
\btheta(\Mvlq)=-\frac{1}{8\pi^2}\keff
\left(\frac{\Mvlq}{\Lcp}\right)^2\log\frac{\Mvlq}{\Lcp}
\,,
}
where
\eq{
\label{def:keff}
\keff\equiv \Lcp(M^{-2}\gamma')_i\im\tr\bigg[\frac{{M^B}^\dag}{\Mvlq}G_i\frac{\Lcp}{\Mvlq}\bigg]
}
quantifies the deviation of $\gamma_{ij}$ compared to $O(1)$ and of $G_i$ compared to the scaling behavior
\eq{
\label{Gi:Lcp}
G_i\sim \frac{\Mvlq}{\Lcp}\,,
}
expected from \eqref{MBd:ui} and $M^B\sim \cM^{Bd}\sim \Mvlq$.

\begin{figure}
\includegraphics[scale=.4]{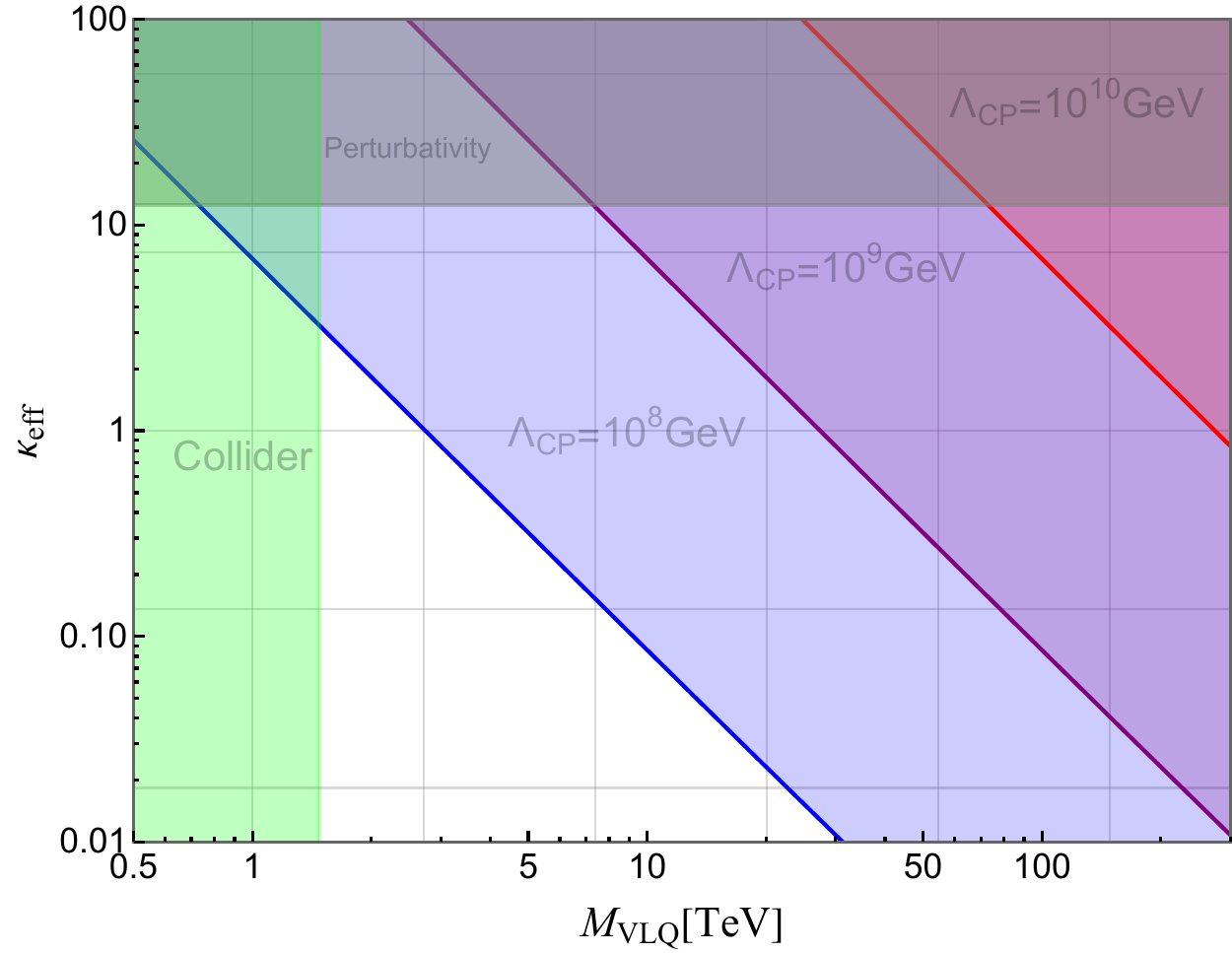}
\caption{\label{fig:keff.MB}%
Limits from $\btheta<10^{-10}$ in the $(\Mvlq,\keff)$ plane following from \eqref{theta:NB:log:MB} with $\keff$
defined in \eqref{def:keff}.
}
\end{figure}

We show in Fig.\,\ref{fig:keff.MB} the possible values for the pair $(\Mvlq,\keff)$ restricted by $\btheta<10^{-10}$
for some values of $\Lcp$.
We can see that for $\Lcp=10^8\,\unit{GeV}$ an order one $\keff$ is perfectly fine for $\Mvlq$ in the TeV scale.
So the suppression of the VLQ Yukawa couplings following from the suppression of the VLQ masses compared to $\Lcp$,
cf.\,\eqref{Gi:Lcp}, is enough to sufficiently suppress $\btheta$ even for order one $\gamma_{ij}$. Consequently,
the necessary suppression of $\btheta$ can be naturally attributed to small Yukawas, which is technically natural,
rather than small Higgs portal couplings $\gamma_{ij}$ to the CP violating sector.

For a general theory of SM augmented by VLQs described by the VLQ-EFT \eqref{lag:sm+vlq:int}, even
if the theory does not follow from a Nelson-Barr theory, 
we can still use our results to track only the contribution from RGE running between $\Mvlq$ 
and the scale $\Lambda$ governing the Wilson coefficients $C_{BdHH}$ and $C_{BBHH}$.
The \emph{additional} contribution at $\Mvlq$ is simply
\eqali{
\label{theta:EFT}
16\pi^2\delta\theta(\Mvlq)=
-
2\Big(\im\tr\left[{Y^{d}}^{-1}Y^B{M^B}^{\dagger}C_{BdHH}\right]
-m_H^2\im\tr\left[{M^B}^{-1}\big(C_{BBHH}-C_{BdHH}{Y^d}^{-1}Y^B\big)\right]\Big)\log\frac{\Mvlq}{\Lambda}
\,.
}
Barring accidental cancellations, this contribution is severely constrained by the experimental limit on 
the electric dipole moment of the neutron.\footnote{%
In general, one needs to add the \emph{tree-level}(-equivalent) contributions
$\theta_d'$ and $\theta_B'$ in eqs.\,\eqref{thetaB'} and \eqref{theta:Mvlq:Md}
which might lead to stronger constraints.
See eq.\,\eqref{ratio:theta} below.
}

Redefining the corresponding dimensionless Wilson coefficients as
\eq{
C_{BdHH}=\frac{c_{BdHH}}{\Lambda}\,,\quad
C_{BBHH}=\frac{c_{BBHH}}{\Lambda}\,,\quad
}
we can rewrite the first two terms in \eqref{theta:EFT} as
\eqali{
\delta\theta_1&=\frac{1}{8\pi^2}c_{1}^{\rm eff}\frac{\Mvlq}{\Lambda}\log\frac{\Lambda}{\Mvlq}
\,,
\cr
\delta\theta_2&=-\frac{1}{8\pi^2}c_{2}^{\rm eff}\frac{m^2_H}{\Mvlq\Lambda}\log\frac{\Lambda}{\Mvlq}
\,,
}
where
\eqali{
\label{def:c1.c2-eff}
c_1^{\rm eff}&\equiv\im\tr[{Y^d}^{-1}Y^B\frac{{M^B}^\dag}{\Mvlq}c_{BdHH}]
\,,
\cr
c_2^{\rm eff}&\equiv\Mvlq\im\tr[{M^B}^{-1}c_{BBHH}]\,.
}
The last contribution in \eqref{theta:EFT} depending on $C_{BdHH}$ is roughly proportional to $\delta\theta_1 
m_H^2/\Mvlq^2$ and is generally subdominant.
We also need to consider the \emph{tree-level}(-equivalent) contributions $\theta_d'$ and $\theta_B'$ in 
eqs.\,\eqref{thetaB'} and \eqref{theta:Mvlq:Md} which do not cancel in VLQ extensions that are not NB.
But they are easily related to the contributions $\delta\theta_1$ and $\delta\theta_2$ by
\eqali{
\label{ratio:theta}
\theta_d'/\delta\theta_1&=
\frac{-4\pi^2v^2\im\tr[{Y^d}^{-1}Y^B {M^B}^{-1} C_{BdHH}]}
{\im\tr[{Y^d}^{-1}Y^B {M^B}^\dag C_{BdHH}]\log(\Lambda/\Mvlq)}
\cr
&\sim \frac{4\pi^2v^2}{\Mvlq^2\log(\Lambda/\Mvlq)}
\approx 2.4\times\frac{1\,\unit{TeV^2}}{\Mvlq^2\log(\Lambda/\Mvlq)}
\,,
\cr
\theta_B'/\delta\theta_2&=
\frac{-2\pi^2}{\lambda\log(\Lambda/\Mvlq)}
\approx -\frac{153}{\log(\Lambda/\Mvlq)}\,.
}
We see that $\theta_B'$ is generally larger than $\delta\theta_2$ by a factor of $10^2$ while the ratio 
$\theta_d'/\delta\theta_1$ may be of order one depending on the scales $\Mvlq$ and $\Lambda$.
So the contraint from $\theta_B'$ should be adapted accordingly from the contraint from $\delta\theta_2$.

\begin{figure}
\includegraphics[scale=.4]{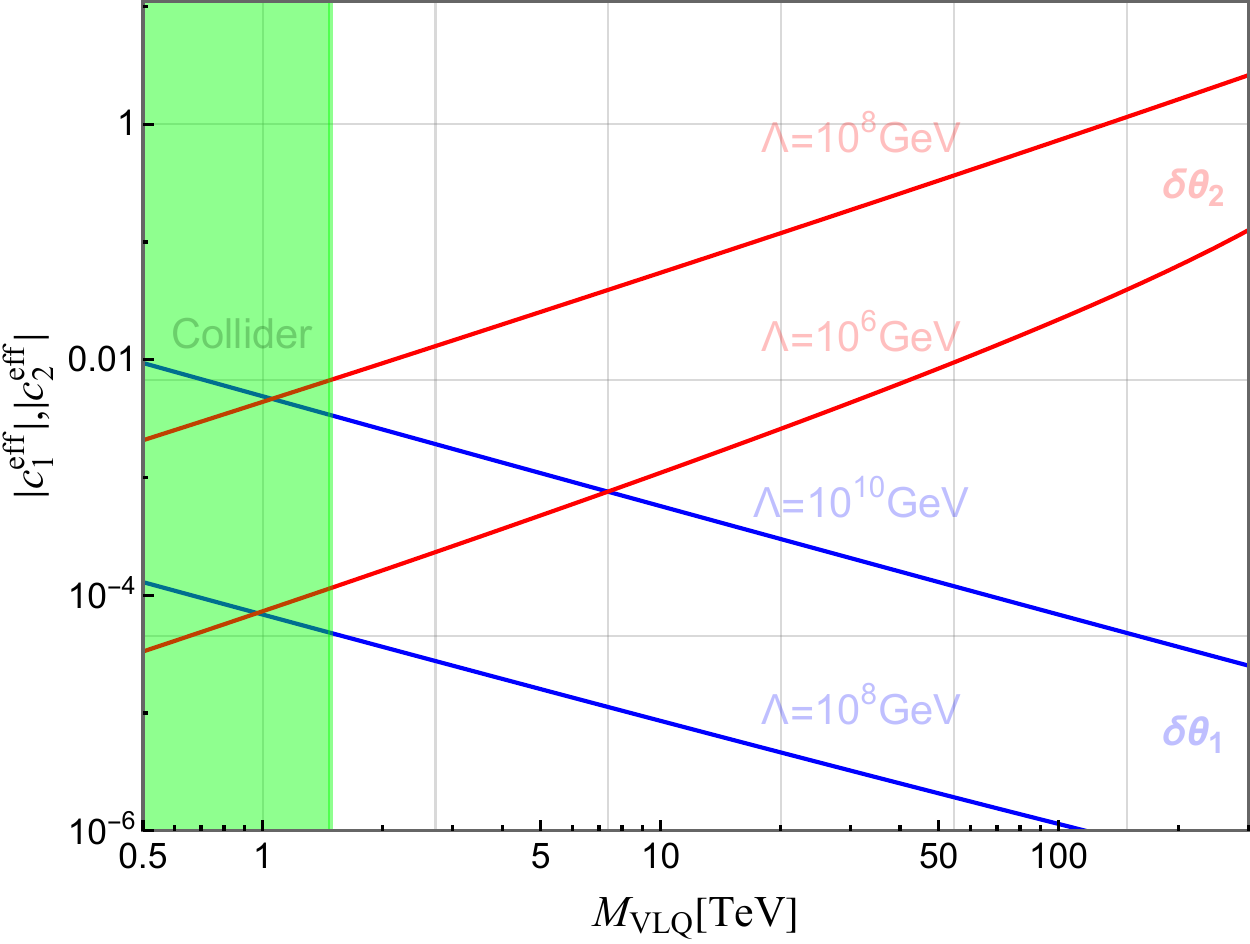}
\caption{\label{fig:c1c2eff.MB}%
Upper limits for $|c_1^{\rm eff}|$ (blue) and $|c_2^{\rm eff}|$ (red)
defined in \eqref{def:c1.c2-eff}
for some scales $\Lambda$
as a function of $\Mvlq$ following from $\delta\btheta<10^{-10}$.
}
\end{figure}

We show in Fig.\,\ref{fig:c1c2eff.MB} the limits on $c_{1}^{\rm eff},c_{2}^{\rm eff}$ following from 
$\delta\btheta<10^{-10}$ assuming the absence of exact cancellations among possible multiple contributions to $\btheta$
in the VLQ-EFT \eqref{lag:sm+vlq:int} or beyond.

Let us start with $c_2^{\rm eff}$.
Excluding severe hierarchies for the VLQs, $c_2^{\rm eff}$ is essentially the imaginary part of the Wilson 
coefficient $C_{BBHH}$ in \eqref{lag:sm+vlq:int} and the EFT expectation is that its dimensionless coefficient 
$c_{BBHH}$ is at most of order one.
Hence, one of the red curves in Fig.\,\ref{fig:c1c2eff.MB} tell us that $|\im c_{BBHH}|\lesssim 0.01$ for a cutoff 
scale of $\Lambda=10^{8}\,\unit{GeV}$ for VLQs of a few TeV.
Taking $\Mvlq=M_{B_1}$ to be the lightest VLQ mass in the basis where $M^B$ is diagonal, we can write this constraint 
more specifically as
\eq{
\Big|\im[c_{BBHH}]^{11}\Big|<4.4\times 10^{-3}\frac{\Mvlq}{1\,\unit{TeV}}\frac{\Lambda}{10^8\,\unit{GeV}}
\bigg[\frac{\log(10^5)}{\log(\Lambda/\Mvlq)}\bigg]\,,
}
where the label $11$ refers to $\bar{B}_{1L}B_{1R}$.
This is quite a strong constraint.

For the effective coupling $c_1^{\rm eff}$, Fig.\,\ref{fig:c1c2eff.MB} tell us that it should be already less than 1\% 
for $\Lambda=10^{10}\,\unit{GeV}$ and $\Mvlq$ of a few TeV.
This constraint may lead to an even stronger constraint on the Wilson coefficient $C_{BdHH}$ 
in case the factor ${Y^d}^{-1}Y^B$ is much larger than unity.
For example, considering the basis of diagonal $Y^d=\hat{Y}^d$ and a Yukawa in the perturbativity limit $Y^B\sim 4\pi$,
we can get
\eq{
\label{pert:Rd}
\frac{4\pi}{y_b}\sim 900\,,\quad
\frac{4\pi}{y_d}\sim 9\times 10^5\,,
}
for the largest and smallest Yukawas.
For up-type VLQs we would have
\eq{
\label{pert:Ru}
\frac{4\pi}{y_t}\sim 14\,,\quad
\frac{4\pi}{y_u}\sim 2\times 10^6\,.
}
For definiteness, we are using the running Yukawas at 1\,\unit{TeV}\,\cite{antusch}.
Assuming the VLQ Yukawas $Y^B$ and the Wilson coefficients $c_{BdHH}$ are not hierarchical, we can write in the $M^B$ 
diagonal basis with $\Mvlq=M_{B_n}$ 
being the largest mass,
\eq{
\label{cBdHH:limit:1}
\Big|\im\Big[Y^B_{1n}[c_{BdHH}]^{n1}\Big]\Big|<9.4\times 10^{-10}\times 
\frac{\Lambda}{10^5\Mvlq}
\bigg[\frac{\log(10^5)}{\log(\Lambda/\Mvlq)}\bigg]\,.
}
This severely limits the imaginary combination of coefficients above.
Even if the imaginary parts of $Y^B_{1n}$ or $[c_{BdHH}]^{n1}$ are suppressed compared to other components with respect 
to $y_d/y_s$, the term associated to $y_s$ still yields
\eq{
\label{cBdHH:limit:2}
\Big|\im\Big[Y^B_{2n}[c_{BdHH}]^{n2}\Big]\Big|<1.9\times 10^{-8}\times 
\frac{\Lambda}{10^5\Mvlq}
\bigg[\frac{\log(10^5)}{\log(\Lambda/\Mvlq)}\bigg]\,.
}
For up-type VLQs, the factor 9.4 in \eqref{cBdHH:limit:1} should be replaced by 4.3 while in \eqref{cBdHH:limit:2}
the factor 1.9 should be replaced by 21.
We conclude that a \emph{generic complex} structure for $Y^B$ and $c_{BdHH}$ is completely excluded for the VLQ-EFT 
\eqref{lag:sm+vlq:int}.

We can again specialize to Nelson-Barr theories to illustrate the automatic suppression factors.
We identify $\Lambda=\Lcp$ and
\eq{
c_1^{\rm eff}=\frac{\Mvlq}{\Lcp}\keff\,,
}
where $\keff$ was defined in \eqref{def:keff}.
This coefficient is automatically suppressed for a large separation between $\Lcp$ and $\Mvlq$.
We can also discuss the factor $(\hY^{d})^{-1}Y^B$ in the case of NB-VLQs.
The possible values for this quantity cannot reach the largest value in \eqref{pert:Rd} (or in \eqref{pert:Ru} for 
up-type) because the various components are correlated and we typically have\,\cite{nb-vlq,nb-vlq:more}
\eqali{
\Big|(\hY^{d})^{-1}Y^B\Big|_{\rm max}\sim 9000\,,
\quad
\Big|(\hY^{u})^{-1}Y^T\Big|_{\rm max}\sim 14000\,.
}

Our constraints above are based on the current limits on the electric dipole moment (EDM) of the 
neutron\,\cite{nEDM:exp}:
\eq{
\label{edm:n:exp}
|d_n| < 3.0\times 10^{-13}\,e\,\unit{fm} \quad(90\%\,\text{CL})\,.
}
The constraint from the ${}^{199}{\rm Hg}$ atom\,\cite{EDM:Hg} is comparable.
Besides the $\btheta$ contribution, the EDM of the neutron (and others) have much more suppressed contributions from 
the EDMs and chromo-electric dipole moments (cEDM) of quarks; we briefly discuss them in Sec.\,\ref{sec:edm}.
Projected future measurements of the EDM of neutrons (nEDM)\,\cite{Balashov:2007zj} and of protons 
(pEDM)\,\cite{Alexander:2022rmq} are expected to improve this bound by about three orders of magnitude.
As the electric dipole moment of the proton $d_p$ receives a contribution from $\btheta$ 
similarly\,\cite{ramsey,cirigliano:cpv-gauge} to the 
neutron's $d_n$, an improved limit on $d_p$ translates into an improved limit on $\btheta$.
This implies that the curves shown in Figs.\,\ref{fig:keff.MB} and \ref{fig:c1c2eff.MB} will be lowered accordingly by 
three orders of magnitude.
These future measurements are based on new techniques:
the experiment \cite{Balashov:2007zj} proposes to measure nEDM using ultra-cold neutrons produced in super-fluid liquid 
helium detecting variations in spin precession under strong electric fields parallel and antiparallel to 
an external magnetic field. The experiment \cite{Alexander:2022rmq} seeks the upper limit to pEDM by using a storage 
ring to observe the spin precession of polarized protons caused by the presence of a radial electric field.

In the following,
we discuss the application of our formulas in two simple versions of the Nelson-Barr scheme:
the Bento-Branco-Parada model\,\cite{BBP} and its extension based on non-conventional CP (CP4)\,\cite{cp4}.
The latter will be treated first due to its simple results on $\btheta$.

\subsection{CP4 model}
\label{sec:cp4}

Here we consider the CP4 model of 
Ref.\,\cite{cp4} with the remarkable property that not only the tree level but the one-loop contribution to $\btheta$ 
vanishes.
The model contains two VLQs $B_1,B_2$ and one complex scalar singlet that can be separated into two scalars $S_1,S_2$. 
The imposed CP symmetry is non-conventional and it is called CP4\,\cite{cp4:3hdm} instead of the usual CP.
On the real scalar fields $S_1$ and $S_2$, the CP4 transformation acts as
\eq{
\label{cp4:S_i}
\cp_4:\quad S_1\to S_2\,,\quad
S_2\to -S_1\,.
}
The transformation is also nontrivial on the VLQs:
\eq{
\label{cp4:B}
\mtrx{B_{1R}\cr B_{2R}}
\to 
i\eps\mtrx{B^{cp}_{1R}\cr B^{cp}_{2R}}
\,,
}
where $\eps$ is the antisymmetric matrix with $\eps_{12}=1$, $\psi^{cp}=i\gamma_0\psi^c$ is the usual CP transformation 
and a similar transformation is valid for $(B_{1L},B_{2L})^\tp$.
On the rest of the SM fields, CP4 acts as usual CP.

The details of the CP breaking sector are reviewed in appendix \ref{ap:CP4}.
Here we only adapt the relevant information to the general notation of Sec.\,\ref{sec:NB}.
The dictionary is given in Table\,\ref{tab:dicitionary} of the appendix.

The CP4 symmetry ensures that only $S_1$ gets nonzero vev, i.e., the vevs $u_i$ in the general notation
\eqref{MBd:ui} is
\eq{
[\aver{S_i}]=
[u_i]=(u_S,0)^\tp\,.
}
After the symmetry breaking, the shifted fields $s_i$ are related by $S_1=s_1+u_S$, $S_2=s_2$.
This vev structure implies that the potential is invariant by the residual accidental $\ZZ_2$ symmetry $s_2\to -s_2$.
In turn, this symmetry ensures that the fields $s_1,s_2$ are already the physical fields with definite mass
and the scalar mass matrix is automatically diagonal as
\eq{
\label{cp4:M}
M^2=\diag(M_1^2,M_2^2)\,.
}
The CP4 symmetry and the vev structure also constrain the Higgs portal couplings in \eqref{VHS:general} to the
structure
\eq{
\label{cp4:gamma'}
[\gamma_{ij}]=\gamma_S\id_2\,,\quad
[\gamma'_{i}]=\gamma_Su_S(1,0)^\tp\,.
}

The Yukawa couplings and the bare mass terms follow our general form in \eqref{lag:UV:cp} but with
$\cF_1,\cF_2$ being $2\times 3$ complex matrices further related by
\eq{
\label{F1.F2}
\cF_2=-i\eps \cF_1^*.
}
The bare mass matrix $\cM^B$ is proportional to the identity $\id_2$ with real positive coefficient while
the mixed bare mass term \eqref{MBd:ui} is simply
\eq{
\cM^{Bd}=\cF_1u_S\,.
}
Note that the residual symmetry of flipping the sign of $s_2$ is only broken by $\cF_2$.

Let us analyze the contributions to $\btheta$ in \eqref{theta:Mvlq:sum} and confirm that these one-loop contributions
vanish in this EFT description.
The dominant contribution is proportional to
\eq{
\label{theta:factor1}
\im\tr[{Y^d}^{-1}Y^B{M^B}^\dag C_{BdHH}]\,,
}
where $C_{BdHH}$ can be read off from Table \ref{tab:op:mass:tree}.
This also covers the one-loop threshold contributions \eqref{thetabar:int.si} and \eqref{theta:CbdHH}.
Generically we can write from \eqref{WR:YdYB},
\eq{
\label{theta:factor1:1}
{Y^d}^{-1}Y^B{M^B}^\dag=(\id_3-ww^\dag)^{-1/2}{\cM^{Bd}}^\dag\,.
}
Now in the CP4 model, \eqref{theta:factor1} vanishes because the specific structure \eqref{cp4:M} and \eqref{cp4:gamma'}
allows us to write
\eq{
\label{CBdHH:cp4}
C_{BdHH}=\frac{\gamma_S}{M^2_1}{\cM^{Bd}}(\id_3-ww^\dag)^{+1/2}\,.
}
We reach the same conclusion by looking at \eqref{theta:NB:log}.
From the latter, it is clear that the vanishing occurs because the CP4 symmetry and its symmetry breaking
pattern prohibit the mixing between the two couplings $\cF_1$ and $\cF_2$ because of the residual symmetry on $s_2$.
In the BBP model we will discuss next, \eqref{CBdHH:cp4} cannot be written and different proportions of $\cF_1$ and
$\cF_2$ contribute leaving an imaginary part.

Now we are only left with the contribution in \eqref{theta:Mvlq:sum} suppressed by $m^2_H$.
The simplest way to check that it vanishes is to use \eqref{Gi:Fi.YB} and the second relation in \eqref{Gi.ui},
which gives
\eq{
{M^B}^{-1}C_{BBHH}=w^\dag w\frac{\gamma_S}{M_S^2}\,.
}
Being hermitean, its trace is real.
This confirms the result that, in the CP4 model, $\btheta$ only receives contribution from
two-loops.
However, the contribution from the dead-duck diagram\,\cite{nelson,dine} ---only additionally suppressed by 
$g_s^2/16\pi^2$ if we ignore the flavor structure--- is also estimated to vanish\,\cite{cp4} and the leading 
nonvanishing contribution is expected to be further suppressed.

For the other effective aspects of the model, we refer to Table~\ref{tab:op:mass:tree} for the matching at $\Lcp$ at 
tree-level and Table~\ref{tab:op:vlq:tree} for the matching at $\Mvlq$ at tree-level.
One needs to use the specific couplings in Table\,\ref{tab:dicitionary}.
The only additional dimension five operators generated at one-loop in the VLQ-EFT are the dipole operators in 
eq.\,\eqref{EFT:dim5:dipole}.
For convenience, we list in Table~\ref{tab:wilson.bbp.mass} of appendix the results of Table~\ref{tab:op:mass:tree}
specific to the CP4 model.
We note that the Wilson coefficient of $\cO_H=(H^\dag H)^3$ vanishes in this model.
This property is similar to the case of a real singlet added to the SM where cubic and quartic scalar
couplings are related appropriately\,\cite{Jiang:2018pbd,Gorbahn:2015gxa}.

\subsection{BBP model}
\label{sec:BBP}

The Bento-Branco-Parada model\,\cite{BBP} is the minimal implementation of the Nelson-Barr scheme with \emph{only one}
VLQ of down type.
Therefore $n=1$ in our formulas and $M^B=M_B$ is just a mass.
The scalar potential of the theory is the renormalizable potential invariant by the symmetries $\ZZ_2$ and CP involving
the SM Higgs and the complex singlet $S=(S_1+iS_2)/\sqrt{2}$ which is $\ZZ_2$ odd; $S_1$ is CP even and $S_2$ is CP odd.

Here we only use the relevant relations and leave the details of the scalar potential for the appendix \ref{ap:BBP}.

For CP breaking it is necessary that both vevs $\aver{S_i}$ be nonzero:
\eq{
[\aver{S_i}]=(u_1,u_2)^\tp\,.
}
In the CP basis, the excitations $s_i=S_i-u_i$ are not the mass definite fields and their mass matrix
\eq{
M^2=\mtrx{M^2_{11} & M^2_{12} \cr M^2_{21} & M^2_{22}}\,,
}
contains nonzero $M^2_{12}=M^2_{21}$ which signals the mixing between opposite CP parities.
The symmetry structure constrain the Higgs portal couplings in \eqref{VHS:general} to the
structure
\eq{
\label{BBP:gamma'}
[\gamma_{ij}]=\diag(\gamma_1,\gamma_2)
\,,\quad
[\gamma'_{i}]=[\gamma_{ij}u_j]=(\gamma_1u_1,\gamma_2u_2)^\tp\,.
}

Due to CP, the Yukawa couplings to $s_i$ in \eqref{lag:UV:cp} have the structure
\eq{
\cF_1=\text{real}\,,\quad
\cF_2=i\times\text{real}\,,
}
while the bare mass $\cM^B$ is just a real positive number.
Differently from the CP4 case, the mixed bare mass term \eqref{MBd:ui}, which is a $1\times 3$ matrix, receives two
contributions
\eq{
\label{cMBd:BBP}
\cM^{Bd}=\cF_1u_1+\cF_2u_2\,,
}
real and imaginary respectively.

In this case all the contributions to $\btheta$ in \eqref{theta:Mvlq:sum} are nonvanishing.
The dominant term is proportional to \eqref{theta:factor1} and one part of it can be simplified as
\eqref{theta:factor1:1} which depends on ${\cM^{Bd}}^\dag$.
Writing the Wilson coefficient $C_{BdHH}$ as
\eq{
C_{BdHH}=\big[(M^{-2}\gamma')_i\cF_i\big](\id_3-ww^\dag)^{+1/2}\,,
}
the trace part is then
\eq{
\label{bbp:theta1:1}
\im\tr\Big\{\big[(M^{-2}\gamma')_i\cF_i\big]{\cM^{Bd}}^\dag\Big\}
\,.
}
This factor is generically nonvanishing
as the part in square bracket can not be arranged to give \eqref{cMBd:BBP}.
We can also write explicitly in the form \eqref{theta:NB:log} which gives
\eq{
\btheta(\Mvlq)=-\frac{1}{8\pi^2}\im\tr\left[\cF_1\cF_2^\dag\right]
(M^{-2}\gamma')_i\eps_{ik}u_k
\log\frac{\Lcp}{\Mvlq}
\,,
}
where $\eps_{ij}$ is the antisymmetric tensor in two dimensions with $\eps_{12}=1$.
This vanishes in the CP4 model because $u_2=0$, $M^2$ is diagonal and $\gamma'_2=0$.

Comparing to \eqref{theta:NB:log:MB}, we can write $\keff$ in different forms:
\eqali{
\keff&=
(M^{-2}\gamma')_i\im\tr\Big[\cF_i{\cM^{Bd}}^\dag\Big]\frac{\Lcp^2}{M_B^2}\,,
\cr
&=
\im\tr\left[\cF_1\cF_2^\dag\right]
\frac{\Lcp^2}{M_B^2}
(M^{-2}\gamma')_i\eps_{ik}u_k\,.
}
Note that $\cF_i\Lcp\sim M_B$ and we could use $\Lcp=\sqrt{u_1^2+u_2^2}$ for definiteness.
From Fig.\,\ref{fig:keff.MB} we see that for $\Lcp\ge 10^8\,\unit{GeV}$ the constraint from $\btheta$ currently allows
an order one $\keff$ comfortably.

We can also check that the contribution proportional to $m_H^2$ in \eqref{theta:Mvlq:sum} depends on the same
nonvanishing factor, i.e.,
\eq{
\im\tr[{M^B}^{-1}C_{BBHH}]=(M^{-2}\gamma')_i\im\tr\big[\cF_i{\cM^{Bd}}^\dag\big]\frac{1}{(\cM^B)^2}\,.
}
This factor is proportional to \eqref{bbp:theta1:1}.

Analogously to the CP4 case, the other effective aspects of the model can be inferred from 
Tables~\ref{tab:op:mass:tree} and \ref{tab:op:vlq:tree}, and eq.\,\eqref{EFT:dim5:dipole}, by using 
the specific couplings in Table\,\ref{tab:dicitionary}.

\subsection{Contributions from quark EDMs and cEDMs}
\label{sec:edm}

Let us now briefly consider the subleading CP violation effects on the neutron EDM which reads\,\cite{snomass22:edm}
\eq{
\label{dn:dipoles}
d_n\approx 
-(0.0015\,e\,\unit{fm})\btheta
-0.2d_u+0.78d_d+0.0027d_s
-0.55\,e\,\tilde{d}_u
-1.1\,e\,\tilde{d}_d
\,.
}
This expression should not be considered as a precise relation as these coefficients are subjected to large 
uncertainties; see e.g. Refs.\,\cite{ramsey,grinstein:alp,edm:lattice}.
The effect proportional to $\btheta$ was considered above. The rest of the terms are proportional to the quark EDMs
$d_q$, $q=u,d,s$, and to the quark chromo-electric dipole moments (cEDMs) $\td_q$, $q=u,d$. 
We are neglecting contributions from additional operators which are less important if coming from Wilson coefficients
in the SMEFT of the same order\,\cite{ramsey}.
The quark EDMs and cEDMs can be defined in the low energy theory as
\eq{
\lag\supset \sum_{q=u,d,s}-i\frac{d_q}{2}\bar{q}\sigma^{\mu\nu}T^A\gamma^5 q F_{\mu\nu}
-i\frac{g_s\td_q}{2}\bar{q}\sigma^{\mu\nu}T^A\gamma^5 q G^A_{\mu\nu}
\,.
}
These low energy operators comes respectively from the SMEFT operators $\mathcal{O}_{dB}$ (or $\cO_{dW}$) and 
$\mathcal{O}_{dG}$ with imaginary Wilson coefficients (the real part gives magnetic 
moments).

Focusing on the down quark, the EDM can thus be written as
\eq{
d_d=-2e\frac{v}{\sqrt{2}}\im [C_{d\gamma}]_{11}
=-(1.13\times 10^{-3}e\,\mathrm{fm})v^2\im [C_{d\gamma}]_{11}
\,,
}
where\footnote{%
Here $B$ refers to the gauge boson and not to the VLQ.
}
\eq{
\im C_{d\gamma}=\im \frac{C_{dB}}{g'}+2I_3^d\im \frac{C_{dW}}{g}\,.
}
The factor $v/\sqrt{2}$ comes from the Higgs vev.
Analogously, its cEDM is related by
\eq{
e\td_d=-2e\frac{v}{\sqrt{2}}\im \frac{[C_{dG}]_{11}}{g_s}
=-(1.13\times 10^{-3}e\,\mathrm{fm})v^2\im \frac{[C_{dG}]_{11}}{g_s}
\,.
}

Our Wilson coefficients for the dipole operators were collected in Table\,\ref{tab:op:vlq:tree}.
We see that for down-type VLQs, only operators with $d_R$ are generated and moreover $C_{dW}=0$ because 
these operators originate from integrating out scalar singlets at $\Lcp$.
Focusing on the Nelson-Barr case we obtain
\eqali{
d_d&=(1.13\times 10^{-3}e\,\mathrm{fm})v^2\im [Y^B{M^B}^{-1}\big({-}\ums{3}\big)C^{\sigma \rm eff}_{Bd}]_{11}\,,
\\
e\td_d&=(1.13\times 10^{-3}e\,\mathrm{fm})v^2\im [Y^B{M^B}^{-1} C^{\sigma \rm eff}_{Bd}]_{11}\,,
}
where we have used \eqref{dipole:eft.eff} and the coefficient $C^{\sigma \rm eff}_{Bd}$ was calculated in 
\eqref{dipole.eff}.
Note that we are neglecting the running between $\Lcp$ and $\Mvlq$, and a possible enhancing log factor may be present.

Using the $d_n$ limit \eqref{edm:n:exp} on \eqref{dn:dipoles}, we obtain for the $d_d$ contribution,
\eq{
\Big|v^2 \im[Y^B{M^B}^{-1}\big({-}\ums{3}\big)C^{\sigma \rm eff}_{Bd}]_{11}\Big|
\lesssim 3.4\times 10^{-10}\,.
}
Considering the proper scaling, ignoring the flavor structure and assuming order one $Y^B$,
one can see that
\eq{
v^2{M^B}^{-1}C^{\sigma \rm eff}_{Bd} \lesssim \frac{1}{16\pi^2}\frac{v^2}{\Lcp^2}
\sim 10^{-14}\bigg(\frac{10^8\,\unit{GeV}}{\Lcp}\bigg)^2\,,
}
starts to be relevant only for $\Lcp\lesssim 10^6\,\unit{GeV}$.
The limit from the cEDM is similar and the case of up-type NB-VLQs is analogous.
We conclude that limits from quark EDMs and cEDMs are not relevant for the Nelson-Barr theory with high $\Lcp$.

\begin{figure}
\raisebox{7em}{{\small (a)}}
\hspace{-2em}
\includegraphics[scale=.75]{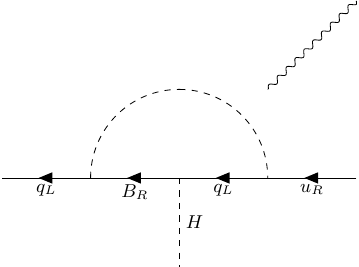}
\hspace{1em}
\raisebox{7em}{{\small (b)}}
\hspace{-2em}
\includegraphics[scale=.75]{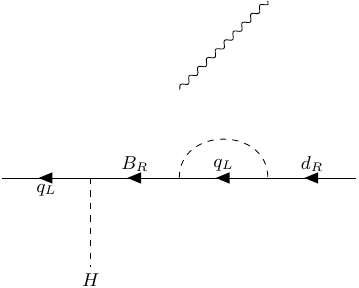}
\caption{\label{diag:dipole}%
One-loop threshold contributions to dipole operators when integrating out VLQs at $\Mvlq$.
}
\end{figure}

We conclude this part by remarking about additional contributions to the dipole operators of the SMEFT that are 
generated in the matching at $\Mvlq$ at one-loop. Being suppressed by $(M^B)^{-2}$, these coefficients are 
parametrically much larger than the one-loop coefficients generated at $\Lcp$.
However, we will show here that their contributions to the quark EDMs and cEDMs vanish.
The additional dipole operators are generated by the diagrams shown in Fig.\,\ref{diag:dipole}.
The wavy line can be a $B_\mu,W_\mu$ or $G_\mu$ that can be attached to many lines.
Some diagrams that are not 1LPI do not contribute in the matching.
To show that these contributions do not enter EDMs or cEDMs, we can focus on 
$\cO_{dB}=\bar{q}_L\sigma^{\mu\nu}d_RHB_{\mu\nu}$. 
In principle all the variants with $W_{\mu\nu}$ and $G_{\mu\nu}$, and changing $d_R\to u_R$, are also generated.
From a spurion analysis, the Wilson coefficient to $\cO_{dB}$ is proportional
to
\eq{
\delta C_{dB}\sim \frac{1}{16\pi^2}G_B Y^d\,,
}
where $G_B$ was defined in \eqref{def:GB}. 
Checking with \texttt{Matchete} \cite{matchete} in fact gives order one prefactor and this is similar to 
the other dipole operators.
Taking the basis where $Y^d=\hY^d$ is diagonal and considering that $G_B$ is hermitean makes
\eq{
\im[\delta C_{dB}]_{11}=0\,.
}
For the contribution to the $u$ quark EDM from $\cO_{uB}$, the coefficient is similarly
\eq{
V\delta C_{uB}\sim \frac{1}{16\pi^2}VG_BV^\dag \hY^u\,,
}
with real (11) or (22) component.
For the latter it is important to remember that we are using the basis where $Y^u=V^\dag\hY^u$, with $V$ being the CKM, 
and that in the broken phase of SMEFT we should consider\,\cite{Jenkins:2017jig}
\eq{
q_L=(V^\dag u_L,d_L)^\tp\,,
}
with $d_L,u_L$ being the mass definite fields.
The fact that there is no CP violating contribution with scale $\Mvlq$ is consistent with the Nelson-Barr construction 
where CP violation is only induced at scale $\Lcp$.

\section{Summary}
\label{sec:summary}

We perform a two stage matching calculation for generic Nelson-Barr theories containing an arbitrary number of CP 
breaking scalars and VLQs at a high scale. 
We integrate out the scalars at the CP breaking scale $\Lcp$ and subsequently the VLQs at $\Mvlq\ll \Lcp$.
The first matching is performed at one-loop up to dimension five operators using the CDE method and the results can be 
applied to generic extensions of the SM with VLQs and scalars interacting through \eqref{lag:UV:cp} 
with a generic scalar potential and not only to Nelson-Barr theories.
Some contributions come from dimension six operators that are transformed to dimension five due to equations of 
motion.
The resulting EFT at this stage is the SM augmented with VLQs and some dimension five operators.
Among these, only the dipole operators arise exclusively at one-loop.
This EFT is then matched onto the SMEFT in the second matching which is performed at tree level and the necessary 
one-loop matching is performed to track the leading one-loop contributions to $\btheta$.
We also obtain the one-loop RGE for $\btheta$ between $\Lcp$ and $\Mvlq$ which is nontrivial and is induced by the 
dimension five operators.
Several contributions cancel within Nelson-Barr theories due to relations between Yukawas \eqref{Gi:Fi.YB} or 
Wilson coefficients \eqref{BBHH->BdHH}.
We confirm that the one-loop running of $\btheta$ also happens within the SMEFT induced by dimension six operators.
In the matching for $\btheta$ in the SMEFT, a subtlety arises due to the additional contribution to the quark masses
from dimension six operators misaligned to the Yukawa couplings.
This leads to a contribution to the QCD $\theta$ term in the matching to the SMEFT.

The results are then applied to generic settings of the NB scheme and it is verified that a large separation between 
$\Lcp$ and $\Mvlq$ already suppresses the one-loop contributions to $\btheta$ sufficiently at the current level.
Our results are equally applicable to generic VLQ extensions of the SM and we find that 
\emph{generic complex} coefficients for the dimension five operator $\bar{B}_Ld_R|H|^2$ is severely constrained 
from its contribution to $\btheta$.
The formulas are also applied to the simplest BBP model\,\cite{BBP} and the CP4 model\,\cite{cp4}. For the latter, we 
confirm the property that all the one-loop contributions to $\btheta$ vanish.
We also analyze the effects of CP violating dipole operators that gives rise to electric dipole moments and 
chromo-electric dipole moments to the quarks.
We confirm that these effects are negligible in the neutron electric dipole moment for a high scale $\Lcp$.
The constraint from $\btheta$ dominates over the ones from EDMs and cEDMs of individual quarks, which is usual 
as for example in the constraints of CP violating interactions between ALPs and quarks\,\cite{grinstein:alp}.

\acknowledgments

G.H.S.A.\ acknowledges financial support by the Coordenação de Aperfeiçoamento de Pessoal de Nível Superior - Brasil 
(CAPES) - Finance Code 001. 
C.C.N.\ acknowledges partial support by Brazilian Fapesp, grant 2014/19164-6, and CNPq, grant 312866/2022-4.

\appendix
\section{Some supertraces}
\label{ap:supertraces}

Here we list the result of some supertraces calculated with the use of the $U$ matrices in \eqref{U:1}, \eqref{U:2} and 
\eqref{U:3} and the classical solutions for $s_i$ in \eqref{si:classical:order}.
Compared to the functional supertrace in \eqref{STr.functional},
the arrow indicates directly the contribution to the effective Lagrangian without the integral $(16\pi^2)^{-1}\int 
d^dx$ 
so that $(16\pi^2)^{-1}$ is also implicit.
We subtract the poles $1/\bar{\eps}$ following the $\overline{\text{MS}}$ scheme and the log $L_{ij}$ is defined in 
\eqref{def:log}.
We are neglecting contributions suppressed by $\mu_H^2/M_i^2$.
\eqali{
&-\ums[i]{2}\Str\Big[K_s^{-1}U_{sH}K_H^{-1}U_{HB}K_B^{-1}U_{Bs}\Big]
-\ums[i]{2}\Str\Big[K_s^{-1}U_{sB}K_B^{-1}U_{BH}K_H^{-1}U_{Hs}\Big]
\hs{.29\linewidth}
\\
&\longrightarrow\ 
(M^{-2}\gamma')_i\Big\{
-\ums{2}\big(\ums[3]{2}\delta_{ij}+L_{ij}\big)(iD_\mu\bar{q}_L)\gamma^\mu HY^BG_j^\dag B_L
-\ums{2}\bar{q}_LHG_i^\dag i\slashed{D}B_L
\cr
&\hspace{6em}
+\big(\delta_{ij}+L_{ij}\big)\bar{q}_LHY^B{M^B}^\dag\big(G_jB_R+F_jd_R\big)
\Big\}
+h.c.
}
\eqali{
&
-\ums[i]{2}\Str\Big[K_s^{-1}U_{sH}K_H^{-1}U_{Hd}K_d^{-1}U_{ds}\Big]
-\ums[i]{2}\Str\Big[K_s^{-1}U_{sd}K_d^{-1}U_{dH}K_H^{-1}U_{Hs}\Big]
\hs{.28\linewidth}
\cr
&\longrightarrow\ 
(M^{-2}\gamma')_i\Big\{
-\ums{2}\big(\ums[3]{2}\delta_{ij}+L_{ij}\big)(iD_\mu\bar{q}_L)\gamma^\mu HY^dF_j^\dag B_L
-\ums{2}\bar{q}_LHF_i^\dag i\slashed{D}B_L
\Big\}
+h.c.
}
\eqali{
&
-\ums[i]{2}\Str\Big[K_s^{-1}U_{sB}K_B^{-1}U_{Bs}\Big]
-\ums[i]{2}\Str\Big[K_s^{-1}U_{sd}K_d^{-1}U_{ds}\Big]
\hs{.42\linewidth}
\cr
&\longrightarrow\ 
\big(\ums{4}\delta_{ij}+\ums{2}L_{ij}\big)\Big\{
\big(\bar{B}_RG_i^\dag +\bar{d}_RF_i^\dag\big)i\slashed{D}\big(G_iB_R +F_id_R\big)
\Big\}
\cr
&\qquad
+\big(\ums{4}\delta_{ij}+\ums{2}L_{ij}\big)\bar{B}_L\big(G_iG_j^\dag+F_iF_j^\dag\big)i\slashed{D}B_L
\cr
&\qquad
+\big(\delta_{ij}+L_{ij}\big)\Big\{\bar{B}_LG_i{M^B}^\dag\big(G_jB_R +F_jd_R\big)+h.c.\Big\}
\cr
&\qquad
+\Big\{
\bar{B}_LG_i{M^B}^\dag\Big[
  M^B{M^B}^\dag M^{-2}_{ik}(\delta_{kj}+L_{kj})
  +\ums{2}\cev{\slashed{D}}\slashed{D}M^{-2}_{ij}
  +\ums{2}M^{-2}_{ik}(\ums[3]{2}\delta_{kj}+L_{kj})\sigma{\cdot}F
\Big]
\cr
&\qquad
    \hs{3em}\times\big(G_jB_R+F_jd_R\big) +h.c.\Big\}
\cr
&\qquad
+\big(\bar{B}_RG_i^\dag+\bar{d}_RF_i^\dag+\bar{B}_LG_i\big)
\Big\{
  -\ums{2}M^{-2}_{ij}M^B{M^B}^\dag i\slashed{D}+\ums{6}M^{-2}_{ij}(i\slashed{D})^3
\cr
&\qquad\hs{3em}
  -\ums{12}M^{-2}_{ij}\big[\sigma{\cdot}Fi\slashed{D}+i\slashed{D}\sigma{\cdot}F\big]
  -\ums{3}M^{-2}_{ik}(\ums[4]{3}\delta_{kj}+L_{kj})\gamma^\nu (D^\mu F_{\mu\nu})
\Big\}(G_jB_R+F_jd_R+G_j^\dag B_L)
\,.
}
\eqali{
-\ums[i]{2}\Str\Big[K_s^{-1}U_{ss}\Big]
\longrightarrow\ 
\ums{2}M^2_{ki}\big(\delta_{ij}+L_{ij}\big)U_{s_js_k}
\,.
\hs{.5\linewidth}
}
\eqali{
-\ums[i]{2}\Str\Big[K_s^{-1}U_{sH}K_H^{-1}U_{Hs}\Big]
&\longrightarrow\ 
\gamma'_i\big(\delta_{ij}+L_{ij}\big)\gamma'_j|H|^2
+\ums{2}\gamma'_iM^{-2}_{ij}\gamma'_j |D_\mu H|^2
\hs{.25\linewidth}
\cr
&\qquad
+(-2)\gamma'_i\big(\delta_{ij}+L_{ij}\big)(M^{-2}\gamma')_j|H|^4
\,.
}
Note that the covariant derivative in $(D^\mu F_{\mu\nu})$, enclosed by parenthesis, only acts on $F_{\mu\nu}$.
This calculation for one VLQ was also checked using \textit{Machete}\,\cite{matchete}.
Clarifications on how to treat open and closed covariant derivatives in different instances can be found in 
appendix B of Ref.\,\cite{cohen:heavy}.

The Table \ref{tab:supertraces.bbp.mass} lists all possible supertraces whose one-loop generated operators have a 
minimum dimension of six. 
Up to dimension 5, most of the operators in this effective theory involving SM fields and VLQs are one-loop corrections 
to operators already generated in tree-level matching. The only exception is the set of operators contributing to the 
dipole terms \eqref{EFT:dim5:dipole} involving VLQs.
But these operators receive contributions from dimension 6 operators which are reduced to dimension 5 after
the use of the EOM \eqref{eom:B}.
This reduction can only happen for dimension 6 operators involving $\slashed{D}B_{L,R}$.
Analyzing the table, few operators involves $\slashed{D}B_{L,R}$.
The only relevant ones are calculated above. The rest only gives $(M^B)^2/M_i^2$ corrections to operators
already appearing in lower order.
So this analysis exhausts all the possible dimension 5 operators generated at one-loop.
\begin{table}[h]
    \centering
\begin{tabular}{|c|c|c|c|}
 \hline
Supertraces & Min dim &Supertraces & Min dim\\
 \hline\rule{0cm}{2em}
$\displaystyle U_{s_is_i}$ 
& 2
& $\displaystyle U_{s_iH}U_{Hs_i}$ 
& 2
\\[1em]
$\displaystyle
 U_{s_iH}U_{HH}U_{Hs_i}$ 
 &
 $\displaystyle
4$
&
$\displaystyle
 \left(U_{s_is_i}\right)^2$ 
 &
 $\displaystyle
4$

\\[1em]
$\displaystyle
 U_{s_is_i}U_{s_iH}U_{Hs_i}$ 
 &
 $\displaystyle
4$

&
$\displaystyle
\left( U_{s_is_i}\right)^3$ 
 &
 $\displaystyle
6$

\\[1em]
$\displaystyle
\left( U_{s_is_i}\right)^2U_{s_iH}U_{Hs_i}$ 
 &
 $\displaystyle
6$

&
$\displaystyle
  U_{s_is_i}U_{s_iH}U_{HH}U_{Hs_i}$ 
 &
 $\displaystyle
6$

\\[1em]
$\displaystyle
 \left(U_{s_iH}U_{Hs_i}\right)^2$ 
 &
 $\displaystyle
4$

&
$\displaystyle
  U_{s_iH}\left(U_{HH}\right)^2 U_{Hs_i}$ 
 &
 $\displaystyle
6$

\\[1em]
$\displaystyle
U_{s_iH}U_{Hf}U_{fH}U_{Hs_i}$ 
 &
 $\displaystyle
5$

&
$\displaystyle
 U_{s_iH}X_{HV}X_{VH}U_{Hs_i}$ 
 &
 $\displaystyle
4$

\\[1em]
$\displaystyle
 U_{s_is_i}\left(U_{s_iH}U_{Hs_i}\right)^2$ 
 &
 $\displaystyle
6$

&
$\displaystyle
 U_{HH}\left(U_{Hs_i}U_{s_iH}\right)^2$ 
 &
 $\displaystyle
6$

\\[1em]
$\displaystyle
 U_{s_iH}U_{Hf_1}U_{f_1f_2}U_{f_2H}U_{Hs_i}$ 
 &
 $\displaystyle
6$

&
$\displaystyle
\left( U_{s_iH}U_{Hs_i}\right)^3$ 
 &
 $\displaystyle
6$

\\[1em]
  $\displaystyle
 U_{s_iB}U_{Bs_i}$ 
 &
 $\displaystyle
3$
&
 $\displaystyle
 U_{s_id}U_{ds_i}$ 
 &
 $\displaystyle
3$
\\[1em]
$ U_{s_iH}U_{HB}U_{Bs_i} $ 
&
$\displaystyle
4$
&
 $U_{s_iH}U_{Hd}U_{ds_i}$
 & 
$\displaystyle
4$
\\[1em]
    $U_{s_iB}U_{Bd}U_{ds_i} $    
&
$\displaystyle
5$
&
$U_{s_iB}U_{BB}U_{Bs_i} $    
&
$\displaystyle
5$
\\[1em]
 $U_{s_iH}U_{Hq}U_{qB}U_{Bs_i} $    
&
$\displaystyle
5$
&
 $U_{s_iH}U_{Hd}U_{dB}U_{Bs_i} $    
&
$\displaystyle
6$
\\[1em]
 $\left( U_{s_iB}U_{Bs_i}\right)^2$
 &
 $\displaystyle
6$
&
  $ \left( U_{s_id}U_{ds_i}\right)^2 $    
  & $\displaystyle
6$
\\[1em]
\hline
\end{tabular}
\caption{All possible supertraces with the SM fields, the VLQ and at least one heavy scalar singlet.
We only show the $U$ matrices and suppress the propagators $K^{-1}$.
}
\label{tab:supertraces.bbp.mass}
\end{table}

\section{Field redefinitions and EOMs}
\label{ap:field.redef}

The use of EOMs to reduce operator redundancy is guaranteed to lead to an equivalent action (same $S$-matrix) 
only when using the EOMs at linear (leading in $1/\Lambda^k$) order; see Ref.\,\cite{Criado:2018sdb} and references 
therein.
The justification relies on the equivalence theorem and field redefinition.
Ref.\,\cite{Criado:2018sdb} also shows a (non-unique) recipe of how to perturbatively remove redundant operators at a 
given order $1/\Lambda^k$:
to remove an operator $f(\bar{\psi})i\slashed{D}\psi$ involving the derivative $i\slashed{D}\psi$ of a fermion $\psi$, 
we need to apply the field redefinition
\eq{
\label{redef.Dpsi}
\bar{\psi}\to \bar{\psi}-f(\bar{\psi})\,,
}
where $f(\bar{\psi})$ is of order $1/\Lambda^k$.
We discuss here some of the necessary field redefinitions to justify the elimination of operators in 
Sec.\,\ref{sec:dim.5}.

We are concerned about one-loop matching at $\Lcp$.
We assume the renormalizable part of the Lagrangian has been properly transformed to canonical form 
\eqref{lag:1-L:redefined} but the effect of the one-loop corrections can be dropped when applying to higher dimensional 
operators and the ``eff'' label will be dropped.
Let us denote this part of the action $\lag_0$ of order $(1/\Lcp)^0$ and analogously $\lag_k$ of order $(1/\Lcp)^k$.

We start with the dimension five operators in \eqref{EFT:s:dim5} and we are interested in eliminating the operators 
involving derivatives.
We focus on the first two operators of the first line which are of order $(16\pi^2\Lcp)^{-1}$ while the rest are of 
order $(16\pi^2)^{-1}\Lcp^{-2}$.
To eliminate the first operator, we apply the field redefinition
\eq{
q_L\to q_L+HC_{DqHB}B_L\,,
}
where the second term is of order $(16\pi^2\Lcp)^{-1}$ and increases the operator dimension by one unit.
This redefinition, when applied to $\lag_0$, successfully eliminates the desired operator and generates operators 
that  would appear with the use of the EOM for $q_L$ but they turn out to be one-loop $(16\pi^2)^{-1}$ suppressed 
contributions with respect to the dimension five operators that already appeared in tree-level matching.\footnote{%
These corrections are of the same order of the corrections induced by the redefinitions \eqref{field.redef:Z} to 
$\lag_k$ with $k\ge 1$.
}
The further terms coming from $\lag_1$ is at least of order $(16\pi^2\Lcp)^{-2}$ which we neglect.
For the second term we analogously apply the redefinition
\eq{
\bar{B}_L\to \bar{B}_L -\bar{q}_LHC_{qHDB}\,.
}
This redefinition also successfully eliminates the desired operator and generates the 
operator that  would appear with the use of the EOMs in \eqref{eom:B},
\eq{
\lag_1\supset \bar{q}_L HC_{qHDB}M^BB_R+h.c.
}
This only induces the correction in \eqref{deltaYB:G'} to $Y^B$.
In this case, however, there are modifications at the next order not captured by the use of EOM\,\cite{Criado:2018sdb}.
The following terms of order $(16\pi^2)^{-1}\Lcp^{-2}$ and dimension six are induced by the dimension five 
operators 
of order $1/\Lcp$ in Table~\ref{tab:op:mass:tree}:
\eq{
\lag_2\supset -\bar{q}_LH|H|^2C_{qHDB}\big[C_{BdHH}d_R+C_{BBHH}B_R\big]+h.c.
}
These are SMEFT operators of the same order as \eqref{EFT:dim6:green} and need to be taken into account if a full 
matching at order $(16\pi^2)^{-1}\Lcp^{-2}$ is pursued.
There are also additional operators of order $(16\pi^2\Lcp)^{-2}$ which we neglect such as the insertions of $C_{qHDB}$
and $C_{qHDB}^\dag$.

For the operators in the second line of \eqref{EFT:s:dim5}, the necessary redefinitions are of order 
$(16\pi^2)^{-1}\Lcp^{-2}$ and only new operators at this order will be generated.
At the same order, appropriate redefinitions can be chosen so as to be equivalent to the use of the EOM of $\lag_0$.
One redefinition that eliminates the second line of \eqref{EFT:s:dim5} is
\eqali{
B_L&\to B_L-\ums{2}i\slashed{D}\big[C_{DBDB}B_R+C_{DBDd}d_R\big]
-\ums{2}C_{DBDB}{M^B}^\dag B_L
\cr
&\quad
-\ums{2}\big[C_{DBDB}{Y^B}^\dag + C_{DBDd}{Y^d}^\dag\big]H^\dag q_L\,,
\cr
\bar{B}_R&\to\bar{B}_R
-\ums{2}\big[\bar{B}_L(-i)\cev{\slashed{D}}+\bar{B}_R{M^B}^\dag\big]C_{DBDB}\,,
\cr
\bar{d}_R&\to\bar{d}_R
-\ums{2}\big[\bar{B}_L(-i)\cev{\slashed{D}}+\bar{B}_R{M^B}^\dag\big]C_{DBDd}\,.
}
This redefinition is obtained by applying \eqref{redef.Dpsi} sequentially until no derivatives are left and we 
choose the form with factors of $1/2$ to remove the two derivatives in a symmetrized way.
This redefinition already removes $i\slashed{D}d_R$.

For the dimension six operators in \eqref{EFT:dim6:green}, the ones with only one derivative can be removed 
straightforwardly by redefinitions in a way equivalent to the use of EOMs.
The same applies to the operator with $\slashed{D}^3$ which has been discussed explicitly in 
Ref.\,\cite{Jenkins:2017dyc}.
For a fermion $\psi$ with mass term $\bar{\psi}_LM\psi_R$, 
to eliminate the operator $\bar{\psi}_R(i\slashed{D})^3C\psi_R$, with $C^\dag=C$, we may use
\eqali{
\psi_R&\to \psi_R-\ums{2}\big[(i\slashed{D})^3-M^\dag M\big]C\psi_R\,,
\cr
\bar{\psi}_L&\to \bar{\psi}_L-\ums{2}\bar{\psi}_R(-i)\slashed{\cev{D}}CM^\dag\,.
}
This explicit redefinition is not written in Ref.\,\cite{Jenkins:2017dyc}.
Here we need three applications of \eqref{redef.Dpsi}.
It is important to emphasize that the necessary field redefinitions are generically non-unique and each choice may 
affect higher order terms differently.

\section{Supertrace for $\mathcal{O}_{H\tilde{G}}$}
\label{ap:UBB}

Here we detail the calculation that leads to the $|H|^2G\tilde{G}$ contribution \eqref{HHGG-tilde} in the one-loop 
matching at $\Mvlq$.
The functional trace we need up to dimension six is
\eqali{\label{trace UBB}
-\ums[i]{2}\Str[K_B^{-1}U_{BB}]&=
\frac{1}{16\pi^2}
\int d^{d} x \tr\Bigg[-M^{3}_B\left(1+\log \frac{\mu}{M^{2}_B}\right) 
\hU_{BB}+\frac{1}{8M_B} \left(\sigma \cdot F\right) \left(\sigma \cdot F\right)\hU_{BB}-\frac{1}{12M_B} F\cdot 
F\hU_{BB}\Bigg]\,,
\cr
&=
\frac{1}{16\pi^2}
\int d^{d} x\bigg\{
\im\tr[{M^B}^{-1}C_{BBHH}]\Big[\frac{g_s^2}{4}G^a_{\mu\nu}G^a_{\alpha\beta}
+\frac{g'^2}{6}B_{\mu\nu}B_{\alpha\beta}
\Big]\eps^{\mu\nu\alpha\beta}|H|^2
\cr
&\qquad
-\re\tr[{M^B}^{-1}C_{BBHH}]
\Big[\frac{g_s^2}{3}G^a_{\mu\nu}G^{a\mu\nu}
+\frac{2g'^2}{9}B_{\mu\nu}B^{\mu\nu}
\Big]|H|^2
\bigg\}\,,
}
where $U_{BB}$ was defined in \eqref{Mvlq:U} and $\hU_{BB}$ is its upper-left fermion-fermion subblock.
In the last line the hypercharge $Y(B_{L,R})=-1/3$ is included.
This, together with $\tr[t_at_b]=\ums{2}\delta_{ab}$ for QCD, explain the relative factor of $2/3=N_c Y^2 1/2$ between 
the terms with $B_{\mu\nu}^2$ and $(G^a_{\mu\nu})^2$.
The result for $|H|^2G\tilde{G}$ was already given in \eqref{HHGG-tilde}.
We also made use of the chiral Fierz identities \cite{Nishi:2004st}:
\eqali{
\tr\left[R\sigma_{\mu\nu}\sigma_{\alpha\beta}\right]&=2\left(g_{\mu\alpha }g_{\nu\beta}-g_{\mu\beta}g_{\nu 
\alpha}+i\epsilon_{\mu\nu \alpha\beta}\right)\,,\\
\tr\left[L\sigma_{\mu\nu}\sigma_{\alpha\beta}\right]&=2\left(g_{\mu\alpha }g_{\nu\beta}-g_{\mu\beta}g_{\nu \alpha}
-i\epsilon_{\mu\nu \alpha\beta}\right)
\,.
}

\section{Details for the CP4 model}
\label{ap:CP4}

We detail here the CP4 model of Ref.\,\cite{cp4} discussed in Sec.\,\ref{sec:cp4}.
The mapping of couplings to the general notation of
Sec.\,\ref{sec:NB} is given in Table\,\ref{tab:dicitionary}.

Considering the nonconventional CP4 transformation \eqref{cp4:S_i}, the scalar potential involving only $S_i$
becomes
\begin{equation}
\label{VSi:cp4}
   V_S(S_1,S_2)=-\frac{1}{2}\mu_{S}\left( S_1^2 +S_2^2\right)+\frac{\lambda_S}{4}\left(S_1^4+
S_2^4\right)+\frac{\lambda_{12}}{2}S_1^2S_2^2,
\end{equation}
while the potential with the Higgs is
\begin{equation}
    V_{SH}(S_1,S_2)=\frac{\gamma_S}{2} (H^\dagger H)\left[S_1^2+S_2^2\right].
\end{equation}

The potential \eqref{VSi:cp4}, induces vevs to $S_i$:
\begin{equation}
\label{cp4:vevs}
\aver{S_1}=u_S=\sqrt{\frac{\mu_S}{\lambda_S}},\quad
\aver{S_2}=0\,,
\end{equation}
where we assume $\la{12}-\la{S}>0$ and the minimum above is connected by symmetry to three other degenerate minima.

Performing a shift, $S_1=s_1+u_S$, $S_2=s_2$,
the potential in terms of $s_1$ and $s_2$ becomes
\eqali{
\label{VS:si:vev}
V_S(s_1,s_2)&=\frac{\lambda_{S}}{4}s_1^4+\frac{\lambda_{S}}{4}s_2^4+u_S\lambda_{S}s^3_1
\\&
+\lambda_{S}u_S^2s_1^2+\frac{\left(\lambda_{12}-\lambda_S\right)u^2_S}{2}s^2_2\\&+\frac{\lambda_{12}}{2}s_1^2s_2^2
+\lambda_{12}u_Ss_2^2s_1
\,,
}
which matches our general form \eqref{VS:general} after dropping the constant terms.
Clearly the mass matrix is already diagonal and given by
\begin{equation}
\label{cp4:M2.lamb}
   M^{2}=\diag(M_1^2,M_2^2)=\begin{pmatrix}2\lambda_{S}u^2_S&0\\
0&\left(\lambda_{12}-\lambda_S\right)u^2_S\end{pmatrix}\,,
\end{equation}
while the couplings $\lambda'_{ijk}$ and $\lambda_{ijkl}$ can be read off.

We can also write the Higgs portal couplings in terms of $s_1$ and $s_2$:
\begin{equation}
\label{VSH:si:vev}
    V_{SH}(s_1,s_2)=\frac{\gamma_S}{2} (H^\dagger H)\left[\left(s_1^2+2u_Ss_1+u_S^2\right)+ s_2^2\right].
\end{equation}
Comparing to our general form \eqref{VHS:general}, we can extract $\gamma_{ij}$ and $\gamma'_i$ which is collected
in Table\,\ref{tab:dicitionary}.

\begin{table}[ht]
\[
\begin{array}{|c|c|c|}
\hline
 \text{Coefficient} &  \text{BBP}&\text{CP4}\\
\hline
M^2_{11} &2\lambda_{11}u^2_1   & 2\lambda_Su^2_S
\\
M^2_{21} & 2\lambda_{12}u_1u_2   & 0
\\
 M^2_{21} & 2\lambda_{12}u_1u_2  & 0
\\
M^2_{22} & 2\lambda_{22}u^2_2  & (\lambda_{12}-\lambda_S)u^2_S
\\
\lambda'_{111} &6u_1\lambda_{11}   & 6u_S\lambda_S
\\
\lambda'_{112} & 2\lambda_{12}u_2   & 0
\\
\lambda'_{121} & 2\lambda_{12}u_2   & 0
\\
\lambda'_{211} & 2\lambda_{12}u_2  & 0
\\
\lambda'_{221} & 2\lambda_{12}u_1  & 2u_S\lambda_{12}
\\
\lambda'_{212} & 2\lambda_{12}u_1  & 2u_S\lambda_{12}
\\
\lambda'_{122} & 2\lambda_{12}u_1   & 2u_S\lambda_{12}
\\
\lambda'_{222} & 6u_2\lambda_{22}  & 0
\\
\lambda_{1111} &6\lambda_{11}   & 6\lambda_{S}
\\
\lambda_{1112} & 0  & 0
\\
\lambda_{1121} & 0  & 0
\\
\lambda_{1211} &  0 & 0
\\
\lambda_{2111} & 0  & 0
\\
\lambda_{1122} &2\lambda_{12}  & 2\lambda_{12}
\\
\lambda_{2211} & 2\lambda_{12}   & 2\lambda_{12}
\\
\lambda_{1212} & 2\lambda_{12}  & 2\lambda_{12}
\\
\lambda_{2121} & 2\lambda_{12}   & 2\lambda_{12}
\\
\lambda_{2112} & 2\lambda_{12}  & 2\lambda_{12}
\\
\lambda_{1221} &2\lambda_{12} &  2\lambda_{12}
\\
\lambda_{2221} & 0  & 0
\\
\lambda_{2212} & 0  & 0
\\
\lambda_{2122} &0  & 0
\\
\lambda_{1222} & 0  & 0
\\
\lambda_{2222} & 6\lambda_{22}  & 6\lambda_{S}
\\
\gamma'_{1} &\gamma_{1}u_{1}   &  \gamma_{S}u_{S}
\\
\gamma'_{2} &\gamma_{2}u_{2}   &  0
\\
\gamma_{11} &\gamma_{1}  &  \gamma_{S}
\\
\gamma_{21} & 0   &  0
\\
\gamma_{22} &\gamma_{2}   &  \gamma_{S}
\\
\cF_{1} & \text{real}  &  \text{--}
\\
\cF_{2} & i\times\text{real}  &
-i\eps \cF_1^*
\\
\hline
\end{array}
\]
\caption{\label{tab:dicitionary}
Coefficients of general form \eqref{VS:general}-\eqref{VHS:general}
for the different potentials: BBP in \eqref{VS:bbp:vev}-\eqref{VSHsi:vev}, CP4 in \eqref{VS:si:vev}-\eqref{VSH:si:vev}.
All the coefficients are symmetric by any permutation of indices.
E.g., $\la{2111}=\la{1211}=\la{1121}=\la{1112}$.
}
\end{table}

\begin{table}[h]
   \centering
\begin{tabular}{|c|c|p{1.7cm}|p{7.2cm}|c|}
\hline
Operator & CP4\\
 \hline\rule{0cm}{2em}
  $\mathcal{O}^{ara'r'}_{BdBd} $
 &
 $\displaystyle
  \frac{[\cF_1]^{ar}[\cF_1]^{a'r'}}{2M_1^2}
  +  \frac{[\cF_2]^{ar}[\cF_2]^{a'r'}}{2M_2^2}
  $
\\[1em]
$\mathcal{O}^{raa'r'}_{dBBd}  $
&
$\displaystyle
 \frac{[\cF^\dag_1]^{ra}[\cF_1]^{a'r'}}{2M_1^2}
 +\frac{[\cF^\dag_2]^{ra}[\cF_2]^{a'r'}}{2M_2^2}
$
\\[1em]
   $\mathcal{O}^{ar}_{BdHH}$
&
$\displaystyle
\frac{[\cF_{1}]^{ar}\gamma_{S}u_S}{M_1^2}$
\\[1em]
$\mathcal{O}_{H \Box}$
 &
$\displaystyle
 -\frac{\gamma^2_{S}}{8\lambda_S^2u^2_S}$
\\[1em]
 $\mathcal{O}_{H }$
 & 0
\\[1em]
 $\displaystyle
 \mathcal{O}_{\delta \lambda_H} $
 &
 $\displaystyle
 \frac{\gamma_{S}^2}{4\la{S}}$
\\[1em]
\hline
\end{tabular}
\caption{Operators and Wilson coefficients for the CP4 model in CP basis.
The convention is the same as of Table\,\ref{tab:op:cp:tree}.
}
\label{tab:wilson.bbp.mass}
\end{table}

\section{Details for the BBP model}
\label{ap:BBP}

Here we detail the BBP model discussed in Sec.\,\ref{sec:BBP}.
The mapping of couplings to the general notation of
Sec.\,\ref{sec:NB} is given in Table\,\ref{tab:dicitionary}.

The potential can be divided into three parts
\begin{equation}
    V=V_H+V_S+V_{HS},
\end{equation}
where $V_H$ is the Higgs potential of the SM. The potential involving only $S$ is
\begin{equation}\label{VSdoBBP}
    V_{S}=S^*S\left[a_1+b_1S^*S\right]+\left(S^2+S^{*^2}\right)\left[a_2+b_2S^*S\right ]+b_3\left(S^4+S^{*^4}\right)\,,
\end{equation}
while the mixed part $V_{HS}$ is described by
\begin{equation}\label{VHSdoBBP}
    V_{H S}= (H^\dagger H)\left[c_1(S^2+S^{*^2})+c_2S^*S\right].
\end{equation}
All these coefficients are real due to
the CP symmetry imposed on the Lagrangian. CP symmetry is then broken spontaneously when the scalar $S$ acquires a
complex VEV.
  \begin{equation}\label{vevsBBP}
   \bra{0}S\ket{0}=Ve^{i\alpha}.
  \end{equation}

Then we can rewrite the potential $V_S$ in \eqref{VSdoBBP} as a function of real $S_1$ and $S_2$ as
\begin{equation}
\label{VSi:bbp}
    V_S(S_1,S_2)=-\frac{1}{2}\mu_{11} S_1^2-\frac{1}{2}\mu_{22} S_2^2+\frac{1}{4} [\lambda_{11}S_1^4+\lambda_{22}
S_2^4+2\lambda_{12}S_1^2S_2^2],
\end{equation}
Analogously, the mixed potential \eqref{VHSdoBBP} becomes
\begin{equation}
\label{VHSi:bbp}
     V_{SH}(S_1,S_2)=\frac{1}{2} (H^\dagger H)\left[\gamma_1S_1^2+\gamma_2 S_2^2\right].
\end{equation}
Because of CP and $\mathbb{Z}_2$, the potential is invariant by the independent sign flips: $S_1\to -S_1$ and $S_2\to
-S_2$.
Both symmetries are spontaneously broken by the vevs
\eq{
\aver{S_1}=u_1\,,\quad
\aver{S_2}=u_2\,.
}
They are given by the solutions of the minimization equations:
\eqali{
\lambda_{11} u_1^2+\lambda_{12}u^2_2&=\mu_{11}\,,
\cr
\lambda_{22} u_2^2+\lambda_{12}u^2_1&=\mu_{22}\,.
}
After the spontaneous breaking of CP and $\ZZ_2$, the potential of the theory as a function of
excitations $s_i=S_i-u_i$ is
\eqali{
\label{VS:bbp:vev}
V_S(s_1,s_2)&=\frac{\lambda_{11}}{4}s_1^4+\frac{\lambda_{22}}{4}s_2^4+\frac{\lambda_{12}}{ 2}s_1^2s_2^2
\cr &+u_1\lambda_{11}s^3_1+u_2\lambda_{22}s_2^3+\lambda_{12}u_2s_1^2s_2+\lambda_{12}u_1s_2^2s_1
\\&+\lambda_{11}u_1^2s_1^2+\lambda_{22}u_2^2s^2_2+2\lambda_{12}u_1u_2s_1s_2\,.
}
We can easily extract the mass matrix
\eq{
\label{M:si:BBP}
   M^2=2\begin{pmatrix}\lambda_{11}u^2_1&\lambda_{12}u_1u_2\\
\lambda_{12}u_1u_2&\lambda_{22}u^2_2\end{pmatrix}\,,
}
whose eigenvalues are
\eqali{
M^2_1&=\lambda_{11}u^2_1+\lambda_{22}u^2_2-\sqrt{\left(\lambda_{11}u^2_1-\lambda_{22}u^2_2\right)^2+\left(2\lambda_{12}
u_1u_2\right)^2}\,,\\
M^2_2&=\lambda_{11}u^2_1+\lambda_{22}u^2_2+\sqrt{\left(\lambda_{11}u^2_1-\lambda_{22}u^2_2\right)^2+\left(2\lambda_{12}
u_1u_2\right)^2}\,.
}

The portal with the SM Higgs can be rewritten as the sum
\begin{equation}
\label{VSHsi:vev}
    V_{SH}=\sum^2_{i=1}\frac{1}{2} (H^\dagger H)\left[\gamma_i\left(s_i^2+2u_is_i+u_i^2\right)\right]\,.
\end{equation}
The term involving $u_i^2$ can be absorbed by the redefinition of the quadratic term $|H|^2$ in the SM Higgs potential.
We show in Table~\ref{tab:dicitionary} the coefficients of the potential in \eqref{VS:bbp:vev}-\eqref{VSHsi:vev} in the
general form \eqref{VS:general}-\eqref{VHS:general}.
Being a spontaneously broken potential, one can see that the cubic $\lambda'_{ijk}$ and the quartic $\lambda_{ijkl}$
couplings are related through the vevs $u_i$.
The same applies to $\gamma_{ij}$ and $\gamma'_i$.


\end{document}